\documentclass[twocolumn,tighten]{aastex631}
\pdfoutput=1
\usepackage[utf8]{inputenc}
\usepackage{graphicx}
\usepackage{natbib}
\usepackage{hyperref}
\usepackage{gensymb}
\graphicspath{ {./} }

\newcommand{\Lprime}{\ensuremath{\mathrm{L}^\prime\;}}
\newcommand{\Mjup}{\ensuremath{\mathrm{M_{jup}}}}

\newcommand{\AU}{\ensuremath{\;\mathrm{AU}}}

\makeatletter
\if@twocolumn
\newcommand{\colwidth}{\columnwidth}
\newcommand{\colwidths}{0.9\columnwidth}
\else
\newcommand{\colwidth}{0.49\textwidth}
\newcommand{\colwidths}{0.49\textwidth}
\fi
\makeatother

\received{2022-08-23}
\revised{2022-10-14}
\accepted{2022-10-23}
\submitjournal{AJ}

\begin{document}

\title{Deep orbital search for additional planets in the HR 8799 system}

\author[0000-0001-5684-4593]{William Thompson}
\affiliation{
    University of Victoria,
    Department of Physics and Astronomy,
    3800 Finnerty Rd,
    Victoria, BC V8P 5C2, Canada}

\author[0000-0002-4164-4182]{Christian Marois}
\affiliation{
    National Research Council of Canada Herzberg,
    5071 West Saanich Rd,
    Victoria, BC, V9E 2E7, Canada}

\author[0000-0001-5173-2947]{Clarissa R. Do \'O}
\affiliation{
    Center for Astrophysics and Space Sciences,
    University of California,
    San Diego, La Jolla, CA 92093, USA
}

\author[0000-0002-9936-6285]{Quinn Konopacky}
\affiliation{
    Center for Astrophysics and Space Sciences,
    University of California,
    San Diego, La Jolla, CA 92093, USA
}

\author[0000-0003-2233-4821]{Jean-Baptiste Ruffio}
\affiliation{
    Department of Astronomy, 
    California Institute of Technology,
    Pasadena, CA 91125, USA
}

\author[0000-0003-0774-6502]{Jason Wang}
\affiliation{
    Department of Astronomy,
    California Institute of Technology,
    Pasadena, CA, USA
}

\author[0000-0001-6098-3924]{Andy J. Skemer}
\affiliation{
    Department of Astronomy and Astrophysics,
    University of California,
    Santa Cruz, Santa Cruz, CA 95064, USA
}

\author[0000-0002-4918-0247]{Robert J. De Rosa}
\affiliation{
    European Southern Observatory,
    Alonso de Córdova 3107, Vitacura, Santiago, Chile
}

\author[0000-0003-1212-7538]{Bruce Macintosh}
\affiliation{
    Kavli Institute for Particle Astrophysics and Cosmology,
    Stanford University,
    Stanford, CA 94305, USA
}

\begin{abstract}
    The HR 8799 system hosts four massive planets orbiting 15 and 80 AU. Studies of the system’s orbital stability and its outer debris disk open the possibility of additional planets, both interior to and exterior to the known system. Reaching a sufficient sensitivity to search for interior planets is very challenging due to the combination of bright quasi static speckle noise close to the stellar diffraction core and relatively fast orbital motion. In this work, we present a deep L-band imaging campaign using NIRC2 at Keck comprising 14 observing sequences. We further re-reduce archival data for a total of 16.75 hours, one of the largest uniform datasets of a single direct imaging target. 
    Using a Bayesian modeling technique for detecting planets in images while compensating for plausible orbital motion, we then present deep limits on the existence of additional planets in the HR 8799 system. The final combination shows a tentative candidate, consistent with $4 - 7 \; \Mjup$ at $4 - 5 \;$AU, detected with an equivalent false alarm probability better than $3\sigma$. This analysis technique is widely applicable to archival data and to new observations from upcoming missions that revisit targets at multiple epochs.

\end{abstract}

\section{Introduction}

The HR 8799 planetary system hosts four giant planets still glowing hot from their recent formation.
It was the first multi-planetary system to be directly imaged \citep{maroisDirectImagingMultiple2008a} in 2008.
Since the subsequent detection of a fourth inner planet in 2010 \citep{maroisImagesFourthPlanet2010}, it became the benchmark system in direct imaging.
Extensive follow up observations of the four planets b, c, d, and e have characterized their orbits, compositions 
\citep{
currieCombinedSubaruVLT2011,
skemerFirstLightLBT2012,
konopackyDetectionCarbonMonoxide2013,
skemerDirectlyImagedLT2014,
currieDEEPTHERMALINFRARED2014,
barmanSimultaneousDetectionWater2015,
wertzVLTSPHERERobust2017,
greenbaumGPISpectraHR2018,
gravitycollaborationFirstDirectDetection2019,
wangChemicalAbundanceHR2020,
wahhajz.SearchFifthPlanet2021,
wangDetectionBulkProperties2021,
sepulvedaDynamicalMassExoplanet2022}, and now masses \cite{brandtFirstDynamicalMass2021}.

Precision astrometric monitoring, orbit fitting, and dynamical modeling have found that the planets follow nearly coplanar orbits with relatively low eccentricity (0 to around 0.25)
\citep{
soummerOrbitalMotionHR2011,
bergforsVLTNACOAstrometry2011,
currieDirectDetectionOrbit2012,
sudolHighmassFourplanetConfigurations2012,
pueyoReconnaissanceHR87992015,
maireLEECHExoplanetImaging2015,
konopackyASTROMETRICMONITORINGHR2016,
zurloFirstLightVLT2016,
wertzVLTSPHERERobust2017,
wangDynamicalConstraintsHR2018, gravitycollaborationFirstDirectDetection2019}.
The orbits appear to form a near-resonant chain, with factor of two period multiples 1b:2c:4d:8e \citep[e.g.][]{maireLEECHExoplanetImaging2015,wangDynamicalConstraintsHR2018,
gozdziewskiExactGeneralizedLaplace2020}.
This is fortunate, since dynamical modeling shows that few configurations exist that are stable over millions of years besides these resonant chains.

These same orbital models show that the system's stability is increasingly tenuous if the inner three planets have masses much above $8\;M\mathrm{_{jup}}$, which matches the masses of approximately $7\;M\mathrm{_{jup}}$ derived through bolometric luminosity and evolution models.
Recently, \citet{brandtFirstDynamicalMass2021} combined previous sets of stable orbits from \citet{wangDynamicalConstraintsHR2018} with careful modeling of the Hipparcos-GAIA proper motion anomaly to estimate the masses of c, d, and e as $9.6^{+1.9}_{-1.8} \; \rm{M_{jup}}$.
This dynamical mass measurement is in slight tension with the results from orbital stability and atmosphere modeling; however, any additional planets in the system would impact this dynamical mass measurement.
In their orbit modeling \citet{gozdziewskiMultipleMeanMotion2014} found some stable configurations that extend the resonant chain down to a fifth inner planet of up to roughly 6 \Mjup~near 7.5 AU (1e:3f) or 9.5 AU (1e:2f), and a larger ``generally stable'' region for test particles below $\sim 6-7$ AU.
Brandt et al do consider such a fifth planet, and place $3\sigma$ detectable mass limits at roughly $5.5\;M\mathrm{_{jup}}$ between 3 and 5 AU,  $6\;M\mathrm{_{jup}}$ between 5 and 7 AU, and $7.5\;M\mathrm{_{jup}}$ near 9 AU, but these limits do not consider how the space of stable orbits used to fit the mass may change by adding a fifth planet.

Many groups have undertaken extensive direct imaging searches for additional planets in the system\citep[e.g.][]{currieDEEPTHERMALINFRARED2014,maireLEECHExoplanetImaging2015,wahhajz.SearchFifthPlanet2021}.
The most sensitive constraints on the mass of an additional inner planet come from \citet{wahhajz.SearchFifthPlanet2021} in YJH (IFS) and K bands (IRDIS).
Using BT-Setl models \citep{spiegelSPECTRALPHOTOMETRICDIAGNOSTICS2012} and assuming an age of 30 Myr, they place $5\sigma$ upper limits of $3.6 \; \mathrm{M_{jup}}$ at 7.5 AU and  $2.8 \; \mathrm{M_{jup}}$ at 9.7 AU.

These limits still leave room in the semi-major axis--mass parameter space where a fifth inner planet could hide; however, there are significant challenges with further improving our sensitivities.
Current observations at such separations are limited by quasi-static speckles \citep{maroisEffectsQuasiStaticAberrations2003}. These speckles produce a non-Gaussian noise distribution that is highly correlated over time and sensitivity improves poorly with increasing integration time.
Observations thus use angular differential imaging \citep[ADI,][]{maroisAngularDifferentialImaging2006a}, at times in combination with spectral differential imaging \citep[SDI,][]{walkerShadesBlackSearching1999,racineSpeckleNoiseDetection1999,maroisEfficientSpeckleNoise2000}, and reference star differential imaging \citep[RDI,][]{wahhajz.SearchFifthPlanet2021}. These greatly improve sensitivity, but require observations to be scheduled near when the system transits the meridian or have a suitable reference star nearby.
Compounding this issue, is orbital motion. Planets at smaller semi-major axes have much shorter orbital periods according to Kepler's third law. 
At a separation of $\sim5$ AU, a planet in this system would move fast enough that observations taken more than a few months apart would start to blur the planet.
All told, this means that considerable integration time is required and that time is challenging to schedule within a few-month window necessary to freeze orbital motion.

Combining images in the presence of orbital motion was previously considered in \citet{malesDirectImagingHabitable2013} and \citet{nowakKStackerKeplerianImage2018}.
These approaches consider ``de-orbitting'' in that images are transformed and stacked to counteract orbital motion.
In doing so, they find that they are able to increase the SNR of a faint candidate despite orbital motion.
One challenge with these approaches is that it becomes difficult to quantify the significance of such a detection in a way that includes uncertainty in the candidate's orbit. A second challenge is that the flux of a candidate can vary freely between epochs even if it is not consistent with later data. This later point may be why \citet{malesDirectImagingHabitable2013} find an increasing false positive rate with increasing orbital motion.

Separately, a large debris disk first noted in the star's spectral energy distribution \citep[SED,][]{sadakaneTwelveAdditionalVegalike1986,zuckermanDustyDebrisDisks2004,rheeCharacterizationDustyDebris2007} and then described by \citet{suDebrisDiskHR2009} lies beyond the known planets and extends outwards to perhaps as far as 1000 AU \citep{matthewsResolvedImagingHR2014}.
The disk is only marginally resolved and models do not yet constrain the inner edge, with estimates varying from $104 ^{+8}_{-12}$ AU \citep{wilnerResolvedMillimeterObservations2018} to 145 AU \citep{boothResolvingPlanetesimalBelt2016} or $170 \pm 40$ AU \citep{faramazDetailedCharacterizationHR2021}.
These works and the additional dynamical studies of \citet{gozdziewskiOrbitalArchitectureDebris2018} and \citet{geilerScatteredDiscHR2019} consider several scenarios for what mechanism may have sculpted the inner edge of the disk, one of which is an additional outer planet between 0.1 \Mjup and 3 \Mjup.
According to \citet{faramazDetailedCharacterizationHR2021} the best limit on an additional outer planet in this regime is $1.25\; \mathrm{M_{jup}}$ by \citet{maireLEECHExoplanetImaging2015}; however, the contrast curves presented in that work end at 70 AU of projected separation.
Further out, \citet{closeSEARCHWIDECOMPANIONS2009} set a lower limit of $\sim 3\; \mathrm{M_{jup}}$ between $\sim200-600\;\mathrm{AU}$,
yet it appears no lower limits have been published on additional outer planets between b and the start of the outer debris disk, or $\sim 100$ to 150 AU.
Though it will not access the inner region of HR 8799 with standard coronagraphic imaging, JWST is poised to place exquisite constraints on outer planets in this regime.

To search for these proposed additional inner and outer planets, we performed an extensive \Lprime imaging campaign at Keck with NIRC2, re-processed archival NIRC2 data using direct $S/N$ optimization, and used a joint Bayesian model of planet orbits and photometry to search for planets despite significant orbital motion.
We present limits on the existence of any additional planets as well as a modest SNR candidate at approximately $4-5 \;\mathrm{AU}$ worthy of further study.

\section{Observations and Processing}

\subsection{Observations}
\begin{deluxetable*}{ccccc}
\tablecaption{Observations\label{tab:obs}}
\tablenum{1}
\tablehead{\colhead{Date} & \colhead{Mask} & \colhead{Integration} & \colhead{FoV Rotation} & \colhead{Seeing: DIMM, MASS, WRF} \\ 
\colhead{(UT)} & \colhead{} & \colhead{(min)} & \colhead{(\degree)} & \colhead{(")}}
\startdata
2009-08-01         & none          & 53                         &  167                            & NA                   \\
2009-10-31         & none          & 34                         &  158                            & 0.6, 0.2, 0.8                     \\
2009-11-01         & none          & 72                         &  162                            & 0.8, 0.2, 1.4                    \\
\hline               
2010-07-21         & corona400     & 26                         &  156                            & 0.8 - 0.5 - NA                   \\
\hline               
2017-07-07         & corona400     & 37                         &  172                            & 0.7, NA,  0.4      \\
2017-07-11         & corona400     & 57                         &  179                            & NA, NA, 0.4     \\
2017-07-12         & corona400     & 90                         &  179                            & NA, NA, 0.4      \\
2017-07-13         & corona400     & 73                         &  178                            & 0.5, 0.4, NA                  \\
2017-07-14         & corona400     & 68                         &  178                            & 0.6, 0.4, NA                   \\
\hline               
2020-08-23         & none          & 62                         &  169                            & 0.6, 0.5, NA       \\
2020-08-25         & none          & 54                         &  165                            & 1.8, NA, NA           \\
2020-08-27         & none          & 43                         &  170                            & 0.8, 0.2, NA     \\
2020-10-07         & none          & 60                         &  170                            & 0.4, 0.2, NA     \\
2020-11-17         & none          & 53                         &  178                            & 0.8, 0.5, NA     \\
\hline               
2021-07-08         & none          & 52                         &  170                            & 0.5, 0.2, NA    \\
2021-07-09         & none          & 65                         &  173                            & 0.5, 0.15, NA    \\
2021-07-10         & none          & 49                         &  173                            & 0.4, 0.2, NA     \\
2021-07-11         & none          & 57                         &  172                            & 0.7, 0.7, NA     \\
\hline
\sidehead{Total}
12 yr baseline     &               & 1005                       &  3069                              &   ~                \\
\enddata
\tablecomments{Observations grouped by year.
        The integration column gives the total science exposure time not including calibrations and overheads.
        Seeing information is summarized from  the Mauna Kea Weather Center Archive where available.
        All sequences were captured with the \Lprime~filter using NIRC2 in narrow mode.
}
\end{deluxetable*}

\begin{deluxetable}{crrrr}
\tablecaption{SNR in combined images\label{tab:reduced}}
\tablenum{2}
\tablehead{\colhead{Year} & \colhead{b} & \colhead{c} & \colhead{d} & \colhead{e}}
\startdata
2009 & 50 & 64 & 33 & 21 \\
2010 & 28 & 48 & 27 & 13 \\
2017 & 92 & 80 & 62 & 22 \\
2020 & 56 & 90 & 75 & 27 \\
2021 & 36 & 62 & 72 & 28 \\
\enddata
\end{deluxetable}

To search for additional planets in the system,
we conducted a campaign of deep $\Lprime$ imaging at Keck using the NIRC2 instrument (PI: K.
Matthews) in 2017, 2020, and 2021.
Our observations for this campaign totalled 14 quarter-nights.
We took observations in pupil tracking mode so that the field of view rotated during each sequence,
but the speckle pattern remained fixed.
This allowed us to process the data with angular differential imaging \citep[ADI,][]{maroisAngularDifferentialImaging2006a} to suppress the halo of quasi-static speckles \citep{maroisEffectsQuasiStaticAberrations2003}.
Each observation was scheduled such that HR 8799 would transit the meridian roughly half-way through the sequence.

We chose $\Lprime$ imaging since it balances the favorable contrast of young planets at longer wavelengths with the need to access tight inner working angles,
as well as limiting noise from the thermal background.

\begin{figure*}[!ht]
    \centering
        \includegraphics[width=\textwidth]{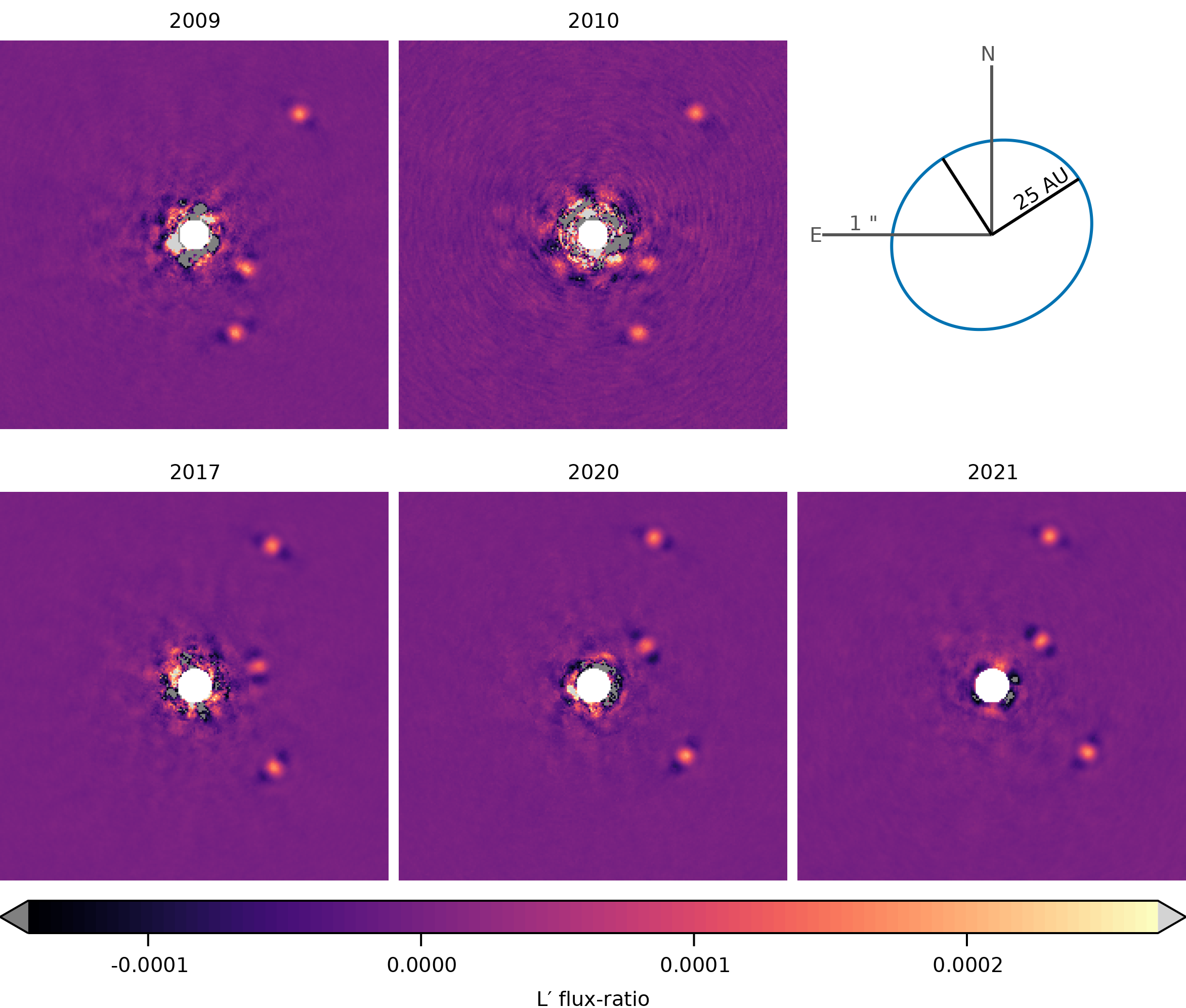}
    
    \caption{
        Combined images from each epoch,
        processed with SNAP
        and cropped to show the inner planets.
        The images are shown on a linear scale with correct throughput,
        normalized using the flux of planet d.
        Values outside the linear colourscale are shown in gray.
        Note that when using SNAP, the forward processed images should not be used for evaluating the noise. Close to the star, the PSFs of any potential planet can produce secondary positive and negative ghosts at other position angles. Instead, we calculate contrast curves using matching backwards-rotated reductions.
        The top panels are comprised of archival data, while the bottom panels are the new campaign.
        Significant orbital motion is visible between epochs.
        The full images including planet b are listed in Appendix \ref{sec:adnl-images}.\label{fig:observations}}
\end{figure*}

\begin{figure}
    \centering
    \includegraphics[width=\colwidth]{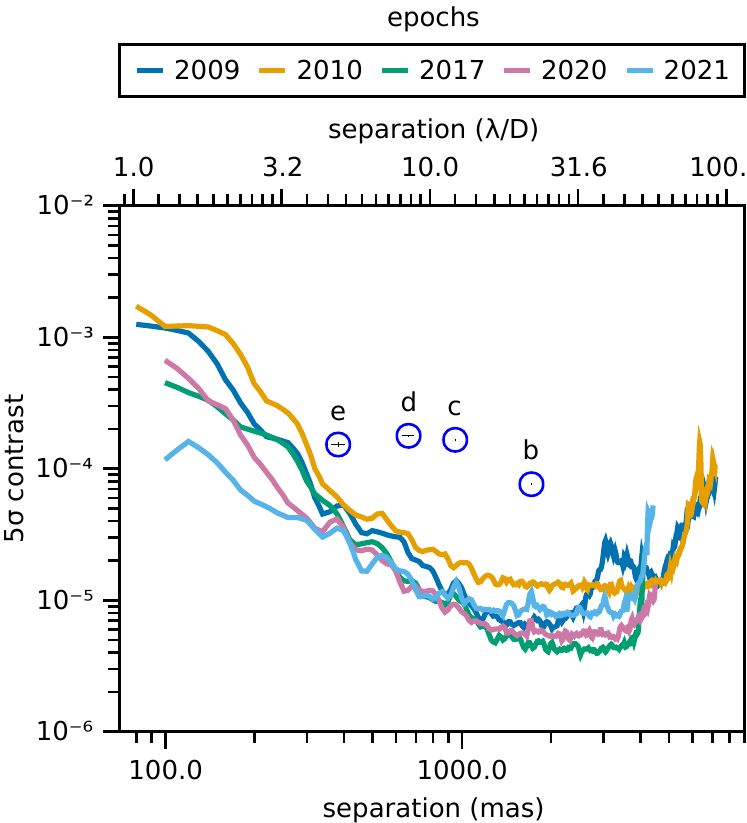}
    \caption{$5\sigma$ contrast limits at each of the five \Lprime epochs on a log-log scale. {Blue circles mark the average separations and photometry from our results}. The contrast achieved close to the star has improved over time, so later epochs have more weight in our modeling.\label{fig:contrast}}
\end{figure}

The observing strategy evolved over the course of the campaign.
The 2017 epoch was captured with the 400 mas diameter Lyot coronagraph,
differential atmospheric refraction (DAR) correction set to acquisition and track,
and included dithering away from transit to improve background subtraction.
For the 2020 epoch, we observed without a coronagraph and with limited dithering.
This is because initial reductions showed that the quasi static speckles near the edge of the mask were not as stable as those in non-coronagraphic datasets 
and because we found that the speckle pattern was not stable between dither positions.
For example, see the $\sim 3.5\times$ improvement in contrast between the otherwise very similar 2017 coronagraphic and 2020/2021  non-coronagraphic epochs near 200-300 mas separation (Figure  \ref{fig:contrast} and Table \ref{tab:obs}).
We did not switch to using the new optical vortex \citep[PI: K. Matthews,][]{serabynKECKOBSERVATORYINFRARED2017} installed in 2015 as it was available only in shared risk mode.
Finally, for the 2021 dataset,
we also chose to set DAR (differential atmospheric refraction) to acquisition only and not acquisition and track to reduce the number of optics moving during our observations.
Our 2021 dataset shown in Figure \ref{fig:observations} achieved the deepest contrast close to the star.
We captured background images at the start and/or ends of each sequence to reduce the thermal background.
The exposure time of each sequence was adjusted to avoid saturating the first Airy ring of the stellar PSF.
Individual exposures were co-added by the detector to create 30-80s exposure images depending on conditions. Higher cadence observations better capture the moment-to-moment variation in the stellar PSF; however, they come at the expense of considerable dead time after each image.
For each sequence, we captured unsaturated non-coronagraphic images at the beginning and ends of each NIRC2 sequence to use as planet PSF templates and contrast calibrations.

\subsection{Archival data selection}\label{sec:obs-selection}
To this campaign,
we added additional data from the Keck archive dating back to some of the first sequences taken of the system.
We considered L' sequences captured by members of our collaboration. 
{
We hoped this data would increase our sensitivity, particularly to a planet whose orbit might have appeared closer to the star during our main campaign.
Combined with our dedicated observing campaign,
this brought the total integration time, not including overheads, bad frames, sky backgrounds, and other calibrations, to 16.75 hours of on-source data at \Lprime.
The sequences taken in 2009 include artifacts at wide separations due to nodding that we exclude from our models of the outer system.}

{Besides this data, the Keck archive contains on the order of a further 8 hours of  L band observations from other researchers. We did not include these sequences due to the large manual effort required to reduce one-off observations captured with varying observing strategies and in some cases, unsuitable choices of focal plane mask that obscure the inner planets.
While not complete, our sample contains on the order of 65\% of all \Lprime data that has been recorded of HR 8799 by NIRC2. If 
we assume that the SNR of the combined observations grows with the square root of the total exposure time, then reducing all  remaining observations could in theory increase the SNR by up to 25\%.}

{Future work could additionally combine data from other wavelengths and observatories. For instance, the Large Binocular Telescope's (LBT) LMIRCam has similar capabilities to NIRC2 at L band.
Observations at other wavelengths e.g. by GPI and SPHERE could also be added, however, using use these observations would require us to assume the color of the planets we search for \textit{a-priori}.}

For a full listing of the sequences used in this work, see Table 1.

\subsection{ADI Reduction}

Angular differential imaging (ADI) is a powerful technique for suppressing quasi-static speckles,
but least-squares based algorithms \citep{lafreniereNewAlgorithmPointSpread2007,maroisGPIPSFSubtraction2014a,soummerDetectionCharacterizationExoplanets2012} suffer from worsening contrast and planet self-subtraction at very small separations.
We therefore developed a new technique for reducing differential imaging sequences called direct $S/N$ optimization \citep{thompsonImprovedContrastImages2021a}.
This technique offers improved contrast close to stars by solving a system of quadratic equations maximizing $S/N$,
rather than linear equations minimizing noise.
Additionally, it optimizes stacks of multiple images in a sequence simultaneously to reduce correlated residual noise.
{See the above reference} for more information and a comparison of direct $S/N$ optimization to LOCI on one of the 2020 sequences presented in this paper.

We used the Signal to Noise Analysis Pipeline (SNAP) to calibrate, align, and reduce each sequence.
We pre-processed all images by subtracting darks, flat fielding with Ks band dome flats, and subtracting sky backgrounds where available.
We applied distortion corrections using the solutions of  \citet{yeldaIMPROVINGGALACTICCENTER2010} and \citet{serviceNewDistortionSolution2016} for data taken after the 2015 NIRC2 servicing.
We then high-pass filtered the images to further suppress thermal background noise using a 25 pixel standard deviation.
This step has the side effect of suppressing any diffuse emission in the system,
though we don't expect to detect any emission from the inner or outer debris disks in these observations.
We aligned the NIRC2 images using an iterative cross-correlation procedure since NIRC2 does not possess ``satellite spots'', off axis faint copies of the stellar PSF used for astrometric and photometric calibration \citep{sivaramakrishnanAstrometryPhotometryCoronagraphs2006,maroisAccurateAstrometryPhotometry2006}.
We aligned each image against the unsaturated PSF template and then stacked to create a master. 
{We then cross-correlated each image against this master to improve their alignment, stacked them to create a new master, and then repeated the procedure a further two times.}
Finally, we rejected bad frames using a correlation threshold (usually 1-5 images per sequence).

We reduced the data using {SNAP} multi-target $S/N$ optimization with batches of 10 images,
and an optimized number of included reference images for each subtraction region.
No parameters of the reduction were changed between sequences to prevent human bias.
The $S/N$ optimization procedure does not include a rejection distance / exclusion zone or other adjustable aggressiveness parameter; all images are used in the optimization including those in which the planet PSF overlaps.

As will be described in the following section, our models assume that the input images have well-calibrated planet throughput. The SNAP pipeline is throughput preserving for point sources as long as the instrument's PSF does not deviate significantly from the unsaturated templates captured before or after the sequence. The $S/N$ optimization algorithm does not produce self-subtraction for the peaks of point sources, and over-subtraction is prevented by the use of optimization, buffer, and subtraction regions. 

For each sequence, we also produced a matching, ``backwards rotated'' image in which the rotation {direction} of the ADI sequence was reversed.
These backwards rotated images have the same noise distribution as the normal images,
but do not contain any significant signal from the planets.
This allows us to calculate a contrast curve for each sequence that is unbiased by planets.
We verified that contrast curves of the backwards rotated sequences matched the regular sequences between planets b, c, d, and e where any additional planets are very unlikely to orbit.

Finally, we combined the processed results of each sequence using a contrast-weighted median stack of images taken less than 3 months apart\footnote{SNAP processes groups of 10 frames {into individual reduced images to reduce correlated noise}. We stacked these individual processed images across nights within the three month window}. During this period, any planet on a circular orbit with semi-major axis greater than or equal to $5 \AU$ would move less than $\frac{1}{3} \; \lambda/D$ . Conveniently, this resulted in {one combined image for each year HR 8799 was observed}.
{The SNRs of planets b, c, d, and e in each of these combined images are presented in Table \ref{tab:reduced}.}

\subsection{Photometric calibration}
Since NIRC2 does not have satellite spots for astrometric and photometric calibration,
we reduced the data in units of contrast relative to the star,
as measured by unsaturated images taken before and/or after the sequence.
However,
variable weather conditions and AO performance during a sequence between the saturated and unsaturated images led to an additional photometric error on the order of 10\% between sequences {which is greater than what we would expect from random variation at the recovered SNR.
To compensate, we measured the flux of planets b, c, d, and e in each epoch and re-scaled the images to the SNR weighted average flux. This correction improved the photometric variation between epochs to the expected level, e.g. a 2\% variation for planet c which is close to what we would expect for a planet with SNR of 60-90.}

To calculate mass we convert from convert from contrast to absolute magnitude using an apparent magnitude of 5.3 for HR 8799 in the \Lprime filter \citep{maroisImagesFourthPlanet2010}.

\begin{figure*}[!ht]
    \centering
    \includegraphics[width=0.8\textwidth]{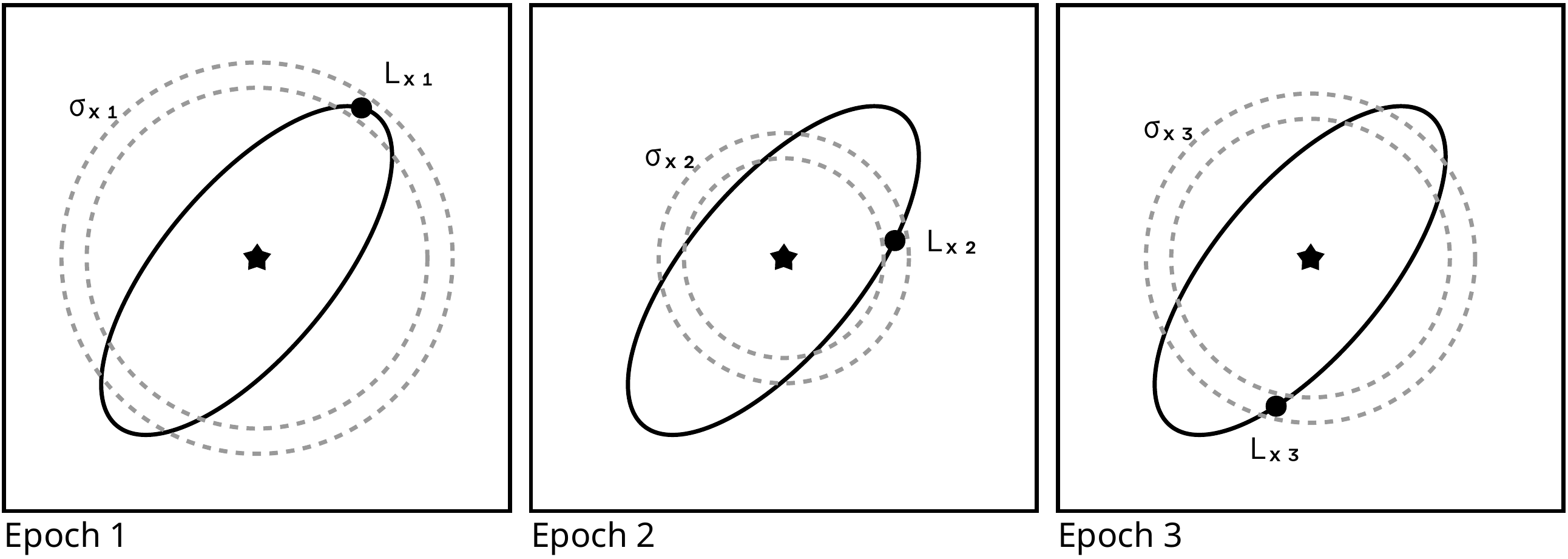}
    \caption{Schematic showing how we model planets across epochs with orbital motion. The solid black line shows the path of a hypothetical planet around a star, with a given set of parameters $(a, i, e, \Omega, \omega, \tau, \mathrm{M}, \Pi)$. We calculate the position $x$ in each image $i$ using Kepler's laws, and measure the flux at that location as $\mathrm{L}_{x,i}$. We then calculate $\sigma_{x,i}$, the uncertainty in $\mathrm{L}_{x,i}$, by measuring the contrast in an annulus at that separation from the matching backwards rotated image.
    $L_{x,i}$ is then compared to the model parameter $L$ which is the same for all epochs using the likelihood function given by equation N.
    \label{fig:schematic}}
\end{figure*}

\section{modeling}\label{sec:modeling}

During the 12 year baseline of our dataset,
any planet with a semi-major axis less than $\approx3500\AU$ would move more than $1\lambda/D$.
This means that the data cannot be na\"ively stacked to improve our sensitivity
as the signal of a planet would not be aligned between epochs.
Since we wish to consider a very large parameter space of inclined and eccentric orbits, simple approaches like rotating and scaling the images would not be effective.

We therefore create a probabilistic model of the system and jointly model the orbits and photometry of the planets.
This combines the process of detecting candidates with orbit fitting \citep[e.g.][]{bluntOrbitizeComprehensiveOrbitfitting2020}
without the intervening step of extracting candidate astrometry at each epoch.
Since astrometry is extracted from images, it follows that if we can model images directly, then extracting astrometry and photometry as intermediate products is not necessary.

Crucially, this allows us to detect the signal of a planet that is too faint to see in a single epoch even if it has moved considerably between images.
In fact, planets with arbitrarily low SNRs per image can grow to detectable levels given a sufficient number of epochs.
The sensitivity and limitations of this method will be expanded upon in a dedicated publication.

\noindent We consider nine parameters:
\begin{itemize}
    \item $M$, the total mass of the system;
    \item $\Pi$, the parallax of the system;
    \item $a$, the semi-major axis;
    \item $e$, the eccentricity; 
    \item $i$, the orbital inclination;
    \item $\Omega$, the longitude of ascending node;
    \item $\omega$, the argument of periapsis; 
    \item $\tau$, the time of periastron passage following the convention of Orbitize! \citep{bluntOrbitizeComprehensiveOrbitfitting2020};
    \item and $L$, the \Lprime flux-ratio.
\end{itemize}
{The physical, geometric, and orbital parameters} define a unique Keplerian orbit through our images and a position at each epoch. We assume that the planet has the same photometry over all observations.
Following \citet{ruffioBayesianFrameworkExoplanet2018} we consider the log-likelihood of a planet having those parameters as 

\begin{equation} \label{eq:1}
\log{\mathcal{L}} \propto 
\sum_i{
    \frac{1}{2\sigma_{x,i}^2}  (L^2 - 2L L_{x,i})
}
\end{equation}
where $i$ is the epoch,
$x$ is the computed position at epoch $i$,
$L_{x,i}$ is the measured photometry extracted at position $x$ from the image $i$,
and $\sigma_{x,i}^2$ is the variance in that photometry.
The schematic in Figure \ref{fig:schematic} illustrates this procedure.

To extract the photometry $L_{x,i}$ efficiently from the images,
{we perform a noise-weighted convolution by an Airy disk  of $1\lambda/D$. Weighing the convolution locally at each pixel by the surrounding contrast is essential so that the peak SNR occurs at the location of the planet, rather than offset in the direction of lower noise. In regions with strongly sloped contrast curves (e.g. planet e) this correction prevents a position bias of up to 30 mas.}
We then look up the photometry at each coordinate using a bi-linear interpolation.
We estimate the variance $\sigma_{x,i}^2$ using the contrast at that separation in each image.
We measured the contrast curves themselves using matching backwards rotated ADI reductions so that the signals of any planets do not bias the contrast.

This likelihood function assumes that our convolved images are maximum likelihood estimates of planet photometry in the presence of Gaussian noise, and that contrast curves extracted from the backwards rotated noise maps provide good estimates of the variance in that estimate at each pixel
\footnote{
One could compute the likelihood by injecting negative fake planets into the raw data; re-performing the post-processing for each epoch, position, and photometry; and examining the residuals;
however, this would increase the compute time by a very large constant factor and is not computationally feasible at this time.}.
\citet{ruffioBayesianFrameworkExoplanet2018} provides a derivation of this likelihood function and shows how non gaussianity does not signficantly effect Bayesian upper limits by that definition.
For detection thresholds on the other hand, section \ref{sec:noise} of this paper discusses how we {correct for} mildly non-Gaussian noise.

This approach is similar to \citet{mawetDeepExplorationEridani2019} in which direct images of $\epsilon$ Eridani are combined with radial velocity data; however, we do not include radial velocities (the planets of HR 8799 orbit almost face on from our perspective) but instead combine images from multiple epochs as in \citet{skemerSiriusImagedMidinfrared2011}.
It also shares some similarities in concept with those of K-stacker \citep{corollerKStackerNewWay2015,nowakKStackerKeplerianImage2018} and for example the recent results of \citet{lecorollerEfficientlyCombiningAlpha2022}.

Note also that we do not use a PCA \citep{soummerDetectionCharacterizationExoplanets2012} or matched filter \citep{ruffioImprovingAssessingPlanet2017} based data processing as in \citet{ruffioBayesianFrameworkExoplanet2018} and \citet{mawetDeepExplorationEridani2019}, neglect the effects of distorted planet PSFs, and simply  {perform a noise-weighted convolution of our data by an Airy disk of $1\lambda/D$}.

Since the purpose of this work is to detect or place limits on the mass of any additional planets rather than precise orbital characterization, we do not consider any systematic errors from the instrument pointing or North angle. Error in registration or North angle could bias the orbital parameters and reduce our ability to recover planets. Thankfully, the results of \citet{yeldaIMPROVINGGALACTICCENTER2010} and \citet{serviceNewDistortionSolution2016} indicate {that the North angle and platescale of NIRC2 are very stable over time, varying less than 0.6\degree and 0.1 $\mathrm{mas/px}$ respectively between 2010 and the service in 2015}.

We consider separate models for the four known planets, an additional outer planet, and an additional inner planet.
Results pertaining to each planet are colored consistently across figures.

The purpose of four known planet models is to confirm we can recover their photometry. These models will additionally confirm that there are no significant North angle offsets between epochs that could impact our ability to detect additional planets. We choose uniform priors for the angular parameters $\omega$ and $\tau$ for each planet and broad but informative priors based on previous work for the orbital plane of the planets, eccentricities, and the planets' L band flux-ratios. {We adopt uniform priors on the planet semi-major axes but truncate them in order to separate the planet models and prevent them from each sampling all four modes of a single posterior}. For the orbital planes, we chose wide Gaussian priors based on previous modeling of the outer planets' orbits and planetary radial velocities by \citet{ruffioRadialVelocityMeasurements2019}. This constrains the direction of motion along our line of sight which roughly halves the size of the parameter space to explore.

\startlongtable
\begin{deluxetable*}{lll}

\tablecaption{Model priors\label{tab:priors}}

\tablenum{3}

\tablehead{\colhead{Parameter} & \colhead{Prior Distribution} & \colhead{Notes}} 

\startdata
$\mathrm{M}$       & $1.52 \pm 0.15 \; \mathrm{M_\odot}$            & \citet{bainesCHARAARRAYANGULAR2012,wangDynamicalConstraintsHR2018,konopackyASTROMETRICMONITORINGHR2016} \\ 
$\Pi$     & $24.46 \pm 0.05\; \mathrm{mas}$           & \citet{gaia-collaborationGaiaEarlyData2021} \\
$i$                & $20.8 \pm 4.5\degree$      & \citet{ruffioRadialVelocityMeasurements2019}\\
$\Omega$           & $89 \pm 27\degree$         & \citet{ruffioRadialVelocityMeasurements2019}\\
$\omega$           & Uniform circular$^*$      &                                             \\ 
$\tau$             & Uniform circular$^*$      &                                             \\
\hline
\sidehead{\textbf{b}}
$L_b$              & $1\times10^{-4} \pm 1\times10^{-4}, L_b \in (0,1)$ & \citet{maroisDirectImagingMultiple2008a} \\
$a_b$              & Uniform(50,180) AU & Images masked outside of 130-200px separation\\ 
$e_b$              & Uniform            & \\
\hline
\sidehead{\textbf{c}}
$L_c$              & $2\times10^{-4} \pm 1\times10^{-4}$ & \citet{maroisDirectImagingMultiple2008a} \\
$a_c$              &  Uniform(30,55) AU  & Images masked outside of 80-120px separation\\
$e_c$              &  Uniform(0,1)       &  \\
\hline
\sidehead{\textbf{d}}
$L_d$              & $1\times10^{-4} \pm 1\times10^{-4}, L_d \in (0,1)$ & \citet{maroisDirectImagingMultiple2008a} \\
$a_d$              & Uniform(20,80) AU  & Images masked outside of 56-80px separation\\
$e_d$              & Uniform(0,1)             & \\
\hline
\sidehead{\textbf{e}}
$L_e$              & $1\times10^{-4} \pm 1\times10^{-4}, L_e \in (0,1)$ & \citet{maroisImagesFourthPlanet2010} \\
$a_e$              & Uniform(8,20) AU    & Images masked outside of 80-120px separation\\
$e_e$              & Uniform(0,1)          &  \\
\hline
\sidehead{\textbf{Outer}}
$L_f$              & Uniform(0, $10^{-5}$)   & \\
$a_f$              & Uniform(100, 160) AU           & Images masked outside of 180-500px separation\\
$e_f$              & Uniform(0,1)              &  \\
\hline
\sidehead{\textbf{Inner}}
$L_f$              & Uniform(0, $10^{-2}$)  & \\
$a_f$              & Uniform(1,14) AU               & Images masked outside of 9-30px separation\\
$e_f$              & Beta(1.1, 5)              & Low-moderate eccentricity\\
\enddata

\tablecomments{$^*$ Parameterized using the arctangent of two standard normal distributions.}
\end{deluxetable*}

We select priors on the flux-ratio that require it to be greater than or equal to zero, but not less than zero. This is because ADI processing introduces dark wings around point sources. If a point source follows a Keplerian orbit through our images, the dark wings will nearly follow this same orbit leading to spurious detections of negative planets and/or reducing the significance of a detection by introducing false uncertainty in its flux-ratio. 
Besides enforcing positivity, we expect the exact shape of priors on the flux-ratio to have little effect on the posterior since we have more than enough data for the likelihood to overwhelm the prior.

For the inner and outer planet models, we again adopt broad but informative priors on the orbital plane of the system and a Beta distribution to prefer low eccentricities. We adopt a uniform distribution for semi-major axis $a$ between 1 and {14 AU which constrains our search to orbits closer in} than planet e.
Finally, for the flux-ratio between the inner planet and star we adopt both a Uniform prior between 0 and $10\times^{-2}$ and a log normal prior centered on the expected pixel values.
Given the amount of data, we expect the posterior to be relatively insensitive to this choice of prior (an assumption we will verify in Section \ref{sec:results-f}).
For the angular parameter $\omega$ which has a uniform prior, we in fact sample from a pair of Gaussian distributions, $\omega_x$ and $\omega_y$ centered at zero, and calculate $\omega=\tan^{-1}(\omega_y, \omega_x)$. This is has the same distribution as a uniform prior on $\omega$, but allows the sampler to easily wrap around past 0 and $2\pi$. We do the same for $\tau$, but restrict it to a domain of $[0,1)$ by diving by $2\pi$.

Finally, for all models we adopted a Gaussian prior on host mass following \citet{konopackyASTROMETRICMONITORINGHR2016} and \citet{wangDynamicalConstraintsHR2018} based on interferometric measurements of the stellar radius by \citet{bainesCHARAARRAYANGULAR2012}.
For parallax $\Pi$ of the system,
we use a tight Gaussian prior from GAIA's EDR3 data release.
We describe the priors further and sources for all parameters in Table \ref{tab:priors}.

\subsection{Detection and Limits}
To evaluate detections, we marginalize over all of the orbital parameters and inspect the flux-ratio ($L$) posterior.
This histogram represents the posterior distribution of the planet's photometry regardless of the orbital parameters. Its central value is the best estimate of the planet's photometry, and the width of the distribution is the uncertainty in that estimate.
A sharp peak that is well-separated from zero indicates a detection.

For ease of comparison with other methods, we summarize this posterior by calculating the SNR as the median divided by half the $84^\mathrm{th} - 17^\mathrm{th}$ percentile distance. This is analogous to the standard SNR calculation used to evaluate point source detections in single images, however it is marginalized over all plausible orbits making it a stricter measure.
In a traditional SNR map each point source is considered separately, even if there were, for example, many significant point sources with varying brightness. 
Here, this SNR is testing the hypothesis that there is a \textit{single} planet with consistent flux.

We can also use a fully Bayesian approach to assessing detections. We can proceed by evaluating the relative probabilities of two models: a model of a planet with a finite brightness ($M_1$) and a model where there is no light from the planet ($M_0$).
The ``Bayes factor'' is then the ratio of the marginal likelihoods of the data given the models times a prior on which model is more likely:
$$
B_{M_1:M_0} = \frac{P(M_1|D)}{P(M_0|D)} \frac{P(M_0)}{P(M_1)}
$$
We adopt the standard prior that both models are \textit{a-priori} equally likely, that is $P(M_1)=P(M_0)$.
The Bayes factor between two models gives the relative probability of $M_1$ compared to $M_0$. For example, if the Bayes factor $B_{M_1:M_0}=10$ then given this data, it is ten times more likely that there is a planet than not.

Often Bayes factors are challenging to calculate numerically since MCMC based methods only produce samples proportional to the posterior density.
However, in our case our two models are said to be ``nested'' since $M_1$ reduces to $M_0$ for $L=0$.
Since our prior on $L$ is uncorrelated with the priors on the orbital parameters, we can calculate the Bayes factor between these nested models using the Savage-Dickey density ratio \citep{dickeyWeightedLikelihoodRatio1971,koopBayesianEconometrics2003}. 
This allows us to calculate the Bayes factor $B_{M_0:M_1}$ as 
$$
B_{M_1:M_0} = \frac{P(L=0)}{P(L=0|D)}
$$
That is, the prior on $L$ evaluated at (or near, for numerical purposes) $L=0$ divided by the marginal posterior of $L$ evaluated at that same location.

A benefit of this approach is that we can {assess} detections without assuming the marginal flux-ratio posterior is Gaussian. 
This could occur even with perfectly Gaussian noise in the images if there is a strong dependence of the flux-ratio on one or more orbital parameters like $a$ or $e$.
Note however that the model $M_1$ itself assumes that the residual noise after post-processing is approximately Gaussian, an assumption we will verify in Section \ref{sec:noise}.
Numerically, this approach requires posterior samples where the flux-ratio is near zero, which for significant detections may be far in the tails of the distribution.
This calculation would require an impractically large number of samples for very significant detections that are well separated from zero. Of course, estimating the Bayes factor accurately in order to asses a highly robust detection is somewhat moot.
In any case, the use of nested sampling may allow one to reliably calculate the Bayes factor between these two models for such significant detections.
For an example of using nested sampling and a Bayes factor to evaluate the presence of an exoplanet from aperture masking interferometry data, see \citet{blakelyTwoRingsMarginally2022}.

A limitation of both the SNR and Bayes factor approaches to evaluating detections is that they consider only a single planet.
If there were, for example, two planets in the data with different orbits and/or flux-ratios to the star, the overall SNR would suffer.

In these instances, the posterior must be examined more closely to disentangle the planets. For this situation, we attempt to reproduce a classical direct imaging SNR map which is not a direct output of this analysis method.
To do so, we first select a given date. Natural choices could be the average date or date of the best input dataset.
At this date, each planet drawn from the posterior has a well defined spatial position which we calculate. 
We then examine the marginal flux-ratio distribution of the posterior draws that fall in each given pixel on that date.
Finally we calculate the SNR of these samples in the same was as above.
These maps are built using the posterior so only include the most \textit{a-posteriori} likely orbit and photometry parameters (no samples are available to perform this analysis along very unlikely orbits). We refer to these as ``photometric accuracy'' maps to distinguish them from traditional SNR maps. 

Finally, we present Bayesian upper limits following \citet{ruffioBayesianFrameworkExoplanet2018} by calculating the $84^\mathrm{th}$ percentile of this same marginal flux-ratio posterior.
In our results, we present the Bayes factor in addition to our analog of the classic SNR whenever this calculation is feasible. The Bayes factor is arguably a more robust quantity, but it lacks a history and established threshold conventions of the SNR in the context of direct imaging.

\subsection{Sampling}

Sampling from images is a difficult problem since the direct imaging likelihood function has {strong modes at the locations of planets and speckles surrounded by large flat regions where the likelihood is negligible}.
Compared to fitting orbits to astrometry points, sampled orbits that fall far away from any modes do not have gradients that pull subsequent samples towards a mode. 
Additionally, fitting near face-on orbits is challenging due to degeneracies between $\Omega$, $\omega$, and $\tau$ in our chosen parameterization.
These effects combine to require small step sizes to explore near the mode and many steps to adequately explore the tails of the posterior.
Our numerical tests showed that {simulated systems with companions injected with an overall SNR of approximately 3-6} are the most computationally demanding to sample, since the sampler must explore multiple peaks (SNRs closer to 1 are associated with posteriors that are broad and relatively smooth and above ~6 the peak dominates).

Knowing this would be computationally demanding, we programmed our model in Julia \citep{bezansonJuliaFastDynamic2012}.
We used forward-mode automatic differentiation \citep{revelsForwardModeAutomaticDifferentiation2016} to calculate the gradients of the log-posterior with respect to each parameter.
This allowed us to use a higher order sampler without manually deriving gradients for each model.
We sampled from the posterior using the No U-Turn Sampler \citep{JMLR:v15:hoffman14a} variant of Hamiltonian Monte Carlo,
as implemented in AdvancedHMC.jl \citep{xuAdvancedHMCJlRobust2020}.
Internally, we used Bijectors.jl \citep{fjeldeFlexibleTransformationsProbability2020} to transform all constrained variables and priors to unconstrained distributions. 

For each model, we used multiple independent chains with a maximum tree-depth of 13 steps.
We initialized each independent chain by drawing 50,000 samples from the priors and picking the sample with the highest posterior density.
After adapting the step size and mass matrix for 30,000 iterations and discarding the first 150,000 iterations, we ran each chain in increments of 150,000 iterations until converged.
We thinned each chain by its auto-correlation time and further discarded occasional chains that failed to adapt and remained at their initial parameters.

We checked for convergence by inspecting trace plots, ensuring all parameters {had effective samples sizes} (ESS) greater than 100, and that the Gelman, Rubin and Brooks diagnostic was less than 1.3 \citep{gelmanInferenceIterativeSimulation1992, brooksGeneralMethodsMonitoring1998}.

\begin{samepage}
    The code used in this paper is available in the Julia packages
    \href{https://sefffal.github.io/PlanetOrbits.jl/dev/}{PlanetOrbits.jl}\footnote{https://sefffal.github.io/PlanetOrbits.jl/dev/} and \href{https://sefffal.github.io/DirectDetections.jl/dev/}{DirectDetections.jl}\footnote{https://sefffal.github.io/DirectDetections.jl/dev/}.
\end{samepage}

\begin{figure*}
    \centering
    \includegraphics[width=\textwidth]{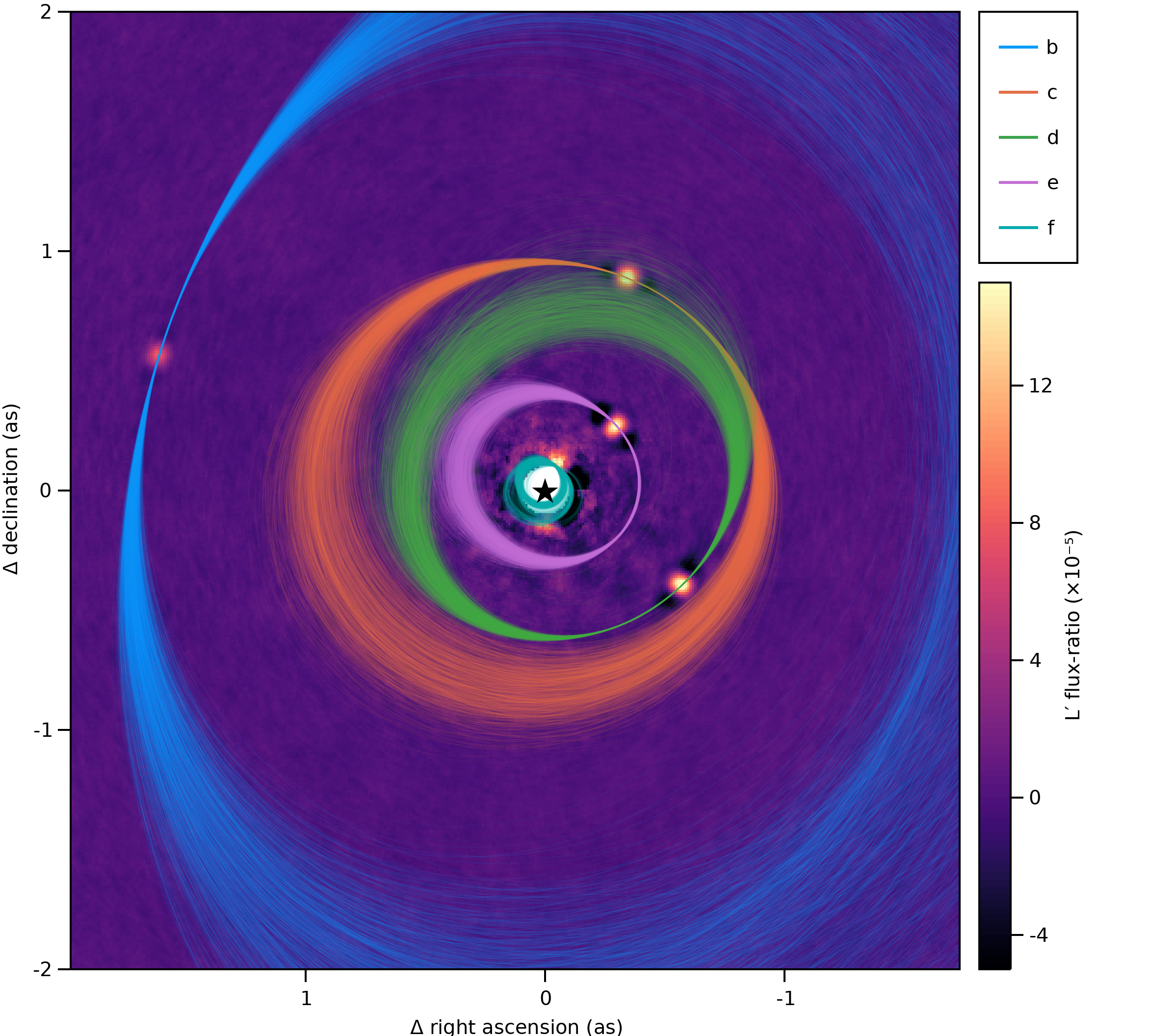}
    \caption{
    Visualization of 1500 orbit draws from the posteriors of the b, c, d, e, and inner planet models over-plotted on the combined 2021 epoch.
    Directly modeling the photometry in our images allows us to simultaneously detect the known planets and produce orbital posteriors that agree with previous fits to extracted astrometry.}\label{fig:bcde-orbits}
\end{figure*}

\subsection{Stability}
We further evaluate the results of our inner planet model by testing them for orbital stability using the Python REBOUND WHFast integrator \citep{2015MNRAS.452..376R} to  integrate sets of 5-planet orbits for 100,000 years.
We determine whether the orbits are stable using the Mean Exponential Growth factor of Nearby Orbits \citep[MEGNO,][]{2003PhyD..182..151C} factor. The orbits that present a MEGNO of $\leq$ 2 for 100,000 years, which would indicate stability up until that time, are then integrated on a range of semi-major axis and eccentricity for 1 Myr to find possibly stable neighboring orbits. \par
We evaluate 5-planet solutions in two ways. First, we start by sampling 5-planet configurations from the posteriors for b, c, d, e and the candidate planet f. Then, we also analyzed the stability of the candidate planet f from our posteriors with the b,c,d and e planet parameters from the $\mathrm{V_d}$ model presented by \citep{gozdziewskiMultipleMeanMotion2014}.

\section{Results}

In this section, we begin by describing the results of our models of the four known planets.
Then, we describe our results of applying the same approach to search for any additional outer and inner planets.

\vspace{2em}
\subsection{Recovery of Known Planets}

\begin{figure*}
    \centering
    \includegraphics[width=0.9\textwidth]{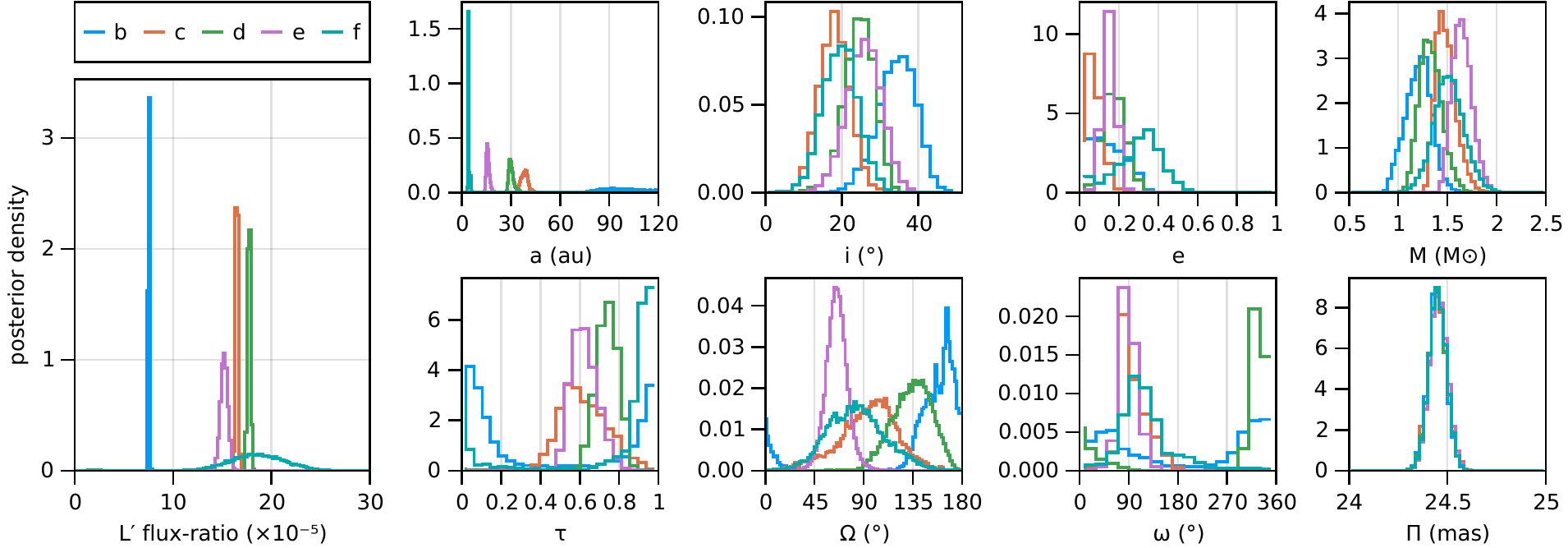}
    \caption{Marginal posteriors of photometry and orbital elements compared between the four known planets and the inner planet model.}\label{fig:model-comparison-all-up}
\end{figure*}

\begin{figure*}
    \centering
    \includegraphics[width=\colwidths]{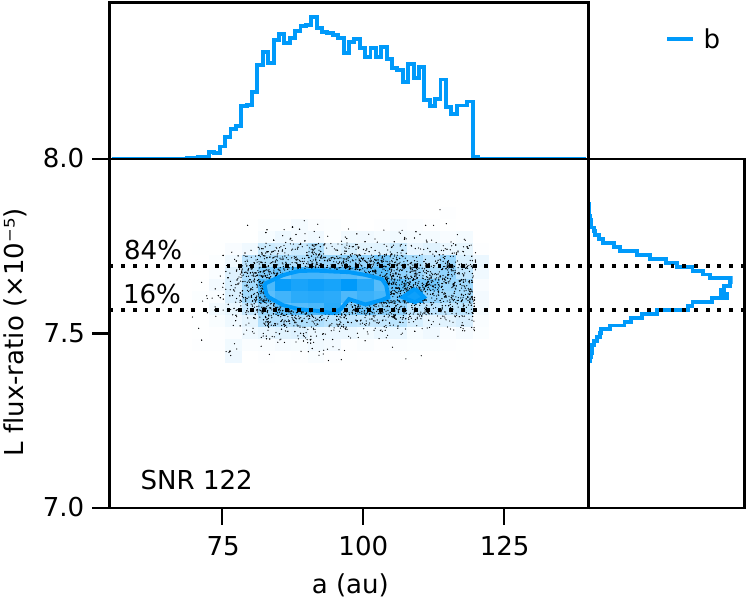}
    \includegraphics[width=\colwidths]{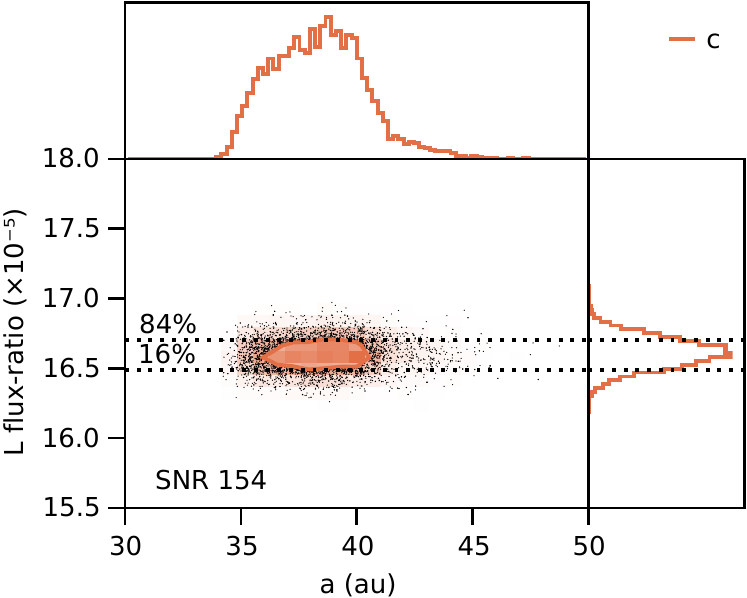}
    \includegraphics[width=\colwidths]{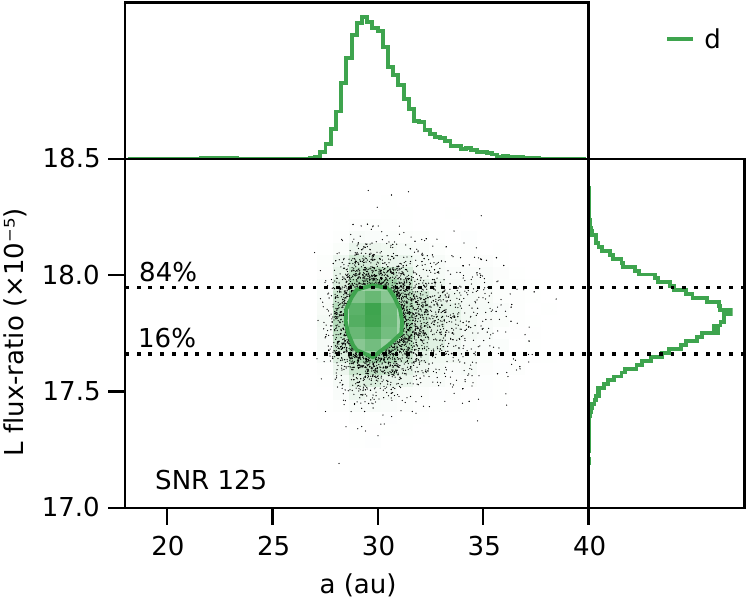}
    \includegraphics[width=\colwidths]{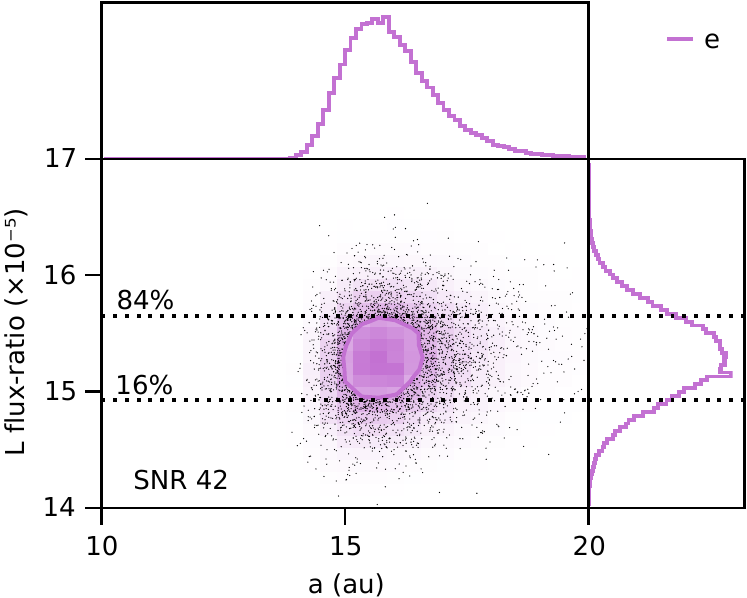}
    \caption{Marginal photometry vs. semi-major axis for planets b, c, d, and e. All four are detected at very high significance and SNRs greater than in any individual epoch. The scales are different between each panel. }\label{fig:esp-phot-vs-a}
\end{figure*}

The four known planets are easily recovered by our models at very high significance.
Figure \ref{fig:bcde-orbits} shows orbital paths drawn from the posterior of the four planet near-resonant model. Figure \ref{fig:model-comparison-all-up} shows the marginal posteriors of the photometry and selected orbital elements for each model.
Despite not extracting astrometry points as an intermediate step, the orbital posteriors are consistent with previous studies \citep{wangDynamicalConstraintsHR2018}.

The marginal $L^\prime$ histograms in Figure \ref{fig:model-comparison-all-up} show that the photometry posteriors are approximately Gaussian distributed and are consistent with previously reported values \citep{maroisDirectImagingMultiple2008a,maroisImagesFourthPlanet2010}. 
The distributions are well separated from zero {which indicate} robust detections. We find SNRs of the planets b, c, d, and e from the combined observations of {122, 154, 125, and 32} respectively.
Compared to the SNR measured at each individual epoch, these SNRs are greater by roughly a {factor of $\sqrt{5}$ which is the ideal improvement in SNR we would expect by combining five observations limited by Gaussian noise.}

Compared to previous studies of the systems orbital configuration using for example GPI \citep{wangDynamicalConstraintsHR2018}, SPHERE \citep{wertzVLTSPHERERobust2017}, and GRAVITY \citep{gravitycollaborationFirstDirectDetection2019}, these longer wavelength observations have reduced astrometric precision. We therefore present our orbital solutions here to show that the orbits derived by directly modeling the photometry in the images are consistent and to build confidence in this approach before applying it to search for additional unseen companions.

\begin{figure}
    \centering
    \includegraphics[width=\colwidth]{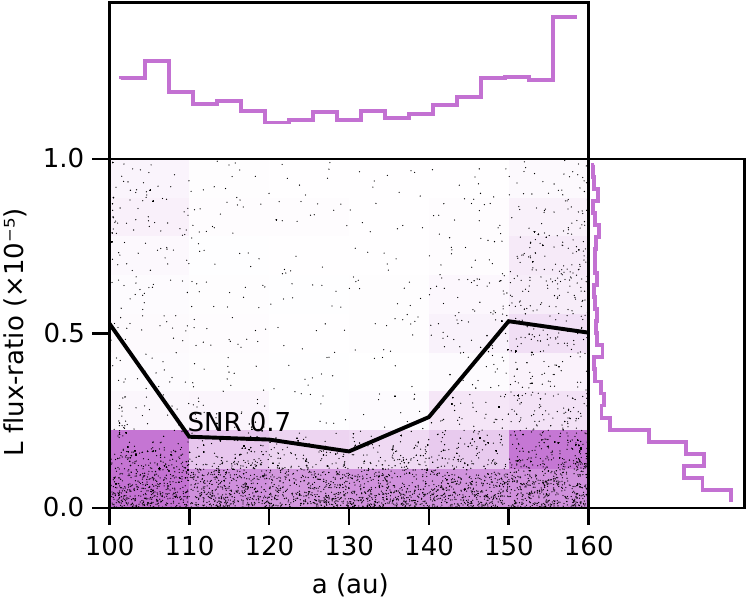} 
    \caption{
    Flux-ratio and semi-major axis marginal posteriors for an additional outer planet between planet b and the start of the outer debris disk.
    {The black line gives the 84th percentile flux as a function of semi-major axis.}
    No planet is detected and 
    our sensitivity is not strongly dependent on any of the orbital parameters within this range of semi-major axis.}
    \label{fig:contrast-out}
\end{figure}

\subsection{Limits on additional outer planets}

Now that we have established that our technique recovers the four known planets,
we turn our search outwards to search for any additional outer planets between 100 and 150 AU.
For this search, we ignored the 2009 epoch due to nodding artifacts {beyond the orbit of planet b}.
{The posterior of this model contained one peak that we identified as a bright artifact on the far top edge of the 2017 epoch. We dropped samples with orbits that intersected that artifact before proceeding with our analysis.}

We find no evidence for a fifth outer planet above an 85th percentile \Lprime contrast of {$4.6\times 10^{-6}$.
{The overall SNR from this posterior is 0.7 and the log Bayes factor for an additional outer planet given this data is -1.6.}
Figure \ref{fig:contrast-out} presents our sensitivity as a function of semi-major axis as well as histogram of the full marginal photometry posterior.
Using the system ages of \citet{sepulvedaDynamicalMassExoplanet2022} and COND models of \citet{baraffeEvolutionaryModelsCool2003}, this corresponds to a $1\sigma$ mass limit of roughly $0.9 \; \mathrm{M_{jup}}$.}

\begin{figure}
    \centering
    \includegraphics[width=\colwidth]{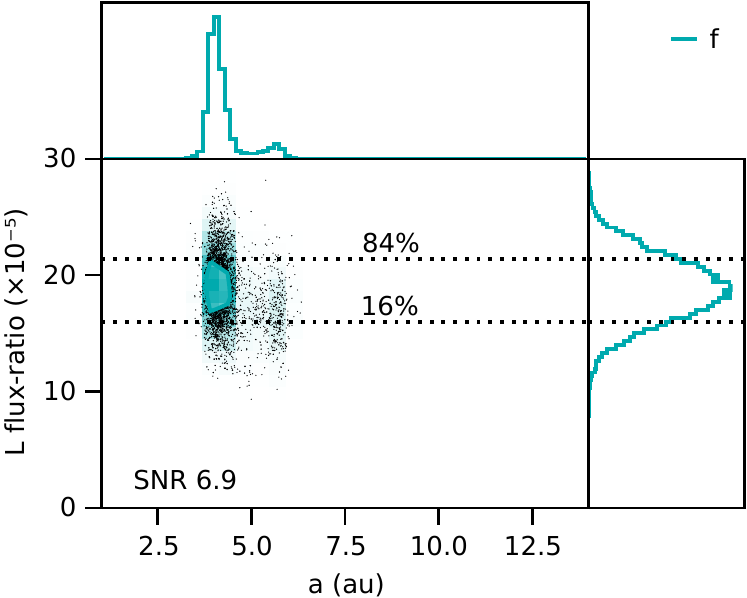}
\caption{ Flux-ratio and semi-major axis posteriors for an additional inner planet with a $1\sigma$ contour over-plotted.
    The dashed lines show 16\% and 84\% percentile limits. The recovered photometry is consistent with that of planets c, d, and e.
    }\label{fig:inn-phot-vs-a}
\end{figure}

\begin{figure*}
    \centering
        \includegraphics[width=\textwidth]{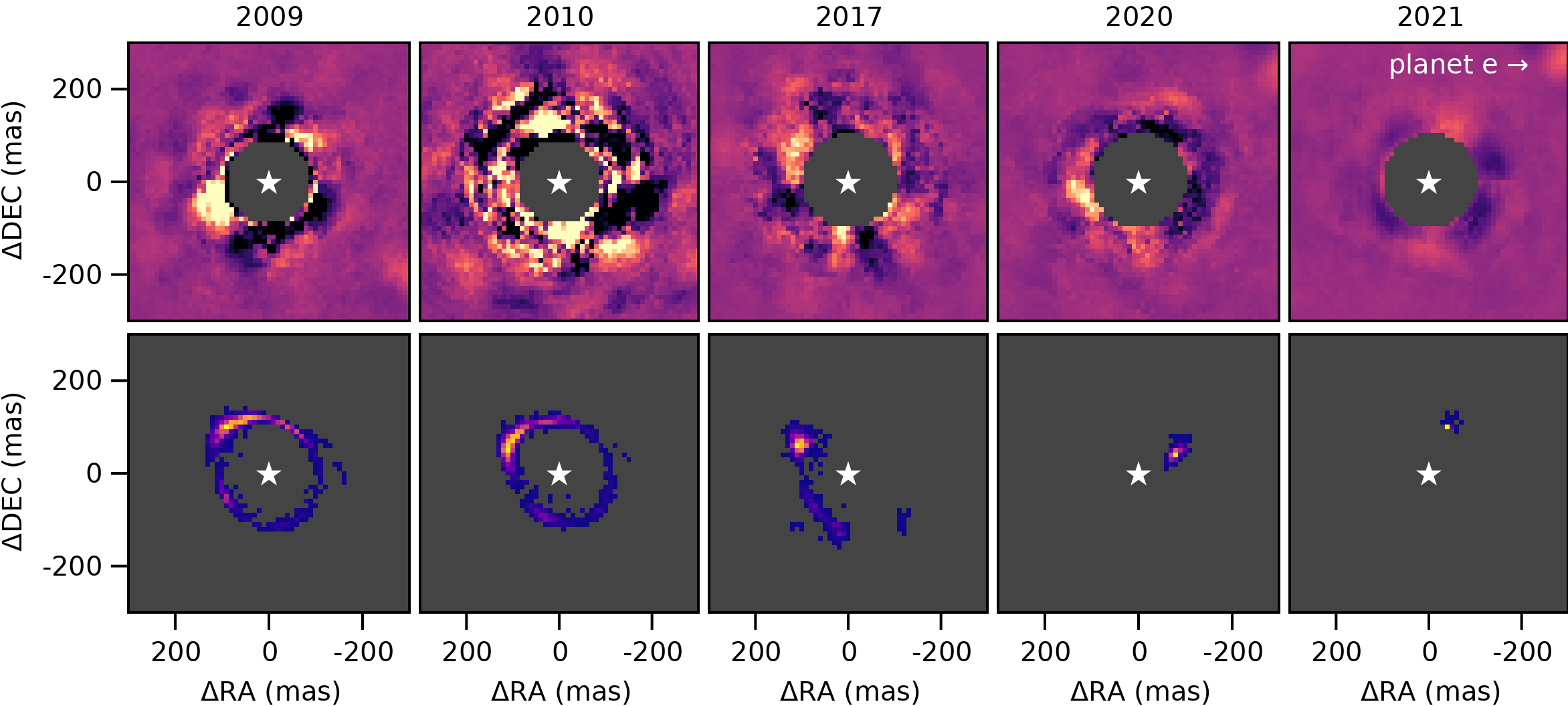}
    \caption{
        \textbf{Top:} {Images} from each of the five epochs. 
        \textbf{Bottom:} 2D marginal position posterior at each epoch. This shows the \textit{a-posteriori} most likely positions for an inner planet at each epoch.
        Note that the posterior is calculated jointly from all images; this figure merely visualizes it at individual epochs.
        Even epochs where the location is poorly constrained can still contribute to the SNR of the model as a whole by reducing the space of plausible orbits and planet photometry.
        }\label{fig:inn-post-combined}
\end{figure*}

\vspace{20pt}
\subsection{Evidence for a fifth inner planet}\label{sec:results-f}

We now consider an additional planet interior to planet e.
Against the full dataset, the model finds a mode close to the star.
Figure \ref{fig:inn-phot-vs-a} shows the marginal photometry vs. semi-major axis posteriors of our single inner planet model.
{The joint posterior of the inner planet model is multi-modal, with 2-3 families of plausible orbits that all pass through the same locations in 2020 and 2021 (Figures \ref{fig:inn-post-combined} and \ref{fig:pos-posts}). This multi-modality is a result of weak photometric constraints in some epochs, leading the model to consider several plausible locations with consistent flux as the peaks found in higher quality epochs. Nonetheless, the marginal photometry posterior is roughly Gaussian and well constrained.}

{
The marginal photometry posterior is well separated from zero,
with a mean that is very similar to the photometry of planets c, d, and e.
We find an SNR of 6.9 and a log Bayes factor of 18.
}
The marginal semi-major axis posterior is {centered at  $4.5$} AU but is cut off below $\sim$4 AU due to detector saturation so we cannot place a firm lower limit. The 85th percentile upper limit is {4.6} AU.
The chains for this posterior are available at 10.5281/zenodo.6823071.

Figure \ref{fig:inn-post-combined} {shows images} from our five epochs with positions calculated from orbits drawn from the posterior.
The model places it NW of the star in 2021 and WNW in 2020. There are three plausible modes in 2017, and the location in earlier epochs is not well constrained.

A full corner plot showing the values, uncertainties, and covariance between all nine parameters is available in Appendix \ref{sec:adnl-images}, Figure \ref{fig:corner-f}.

\begin{figure*}
    \centering
    \includegraphics[width=\textwidth]{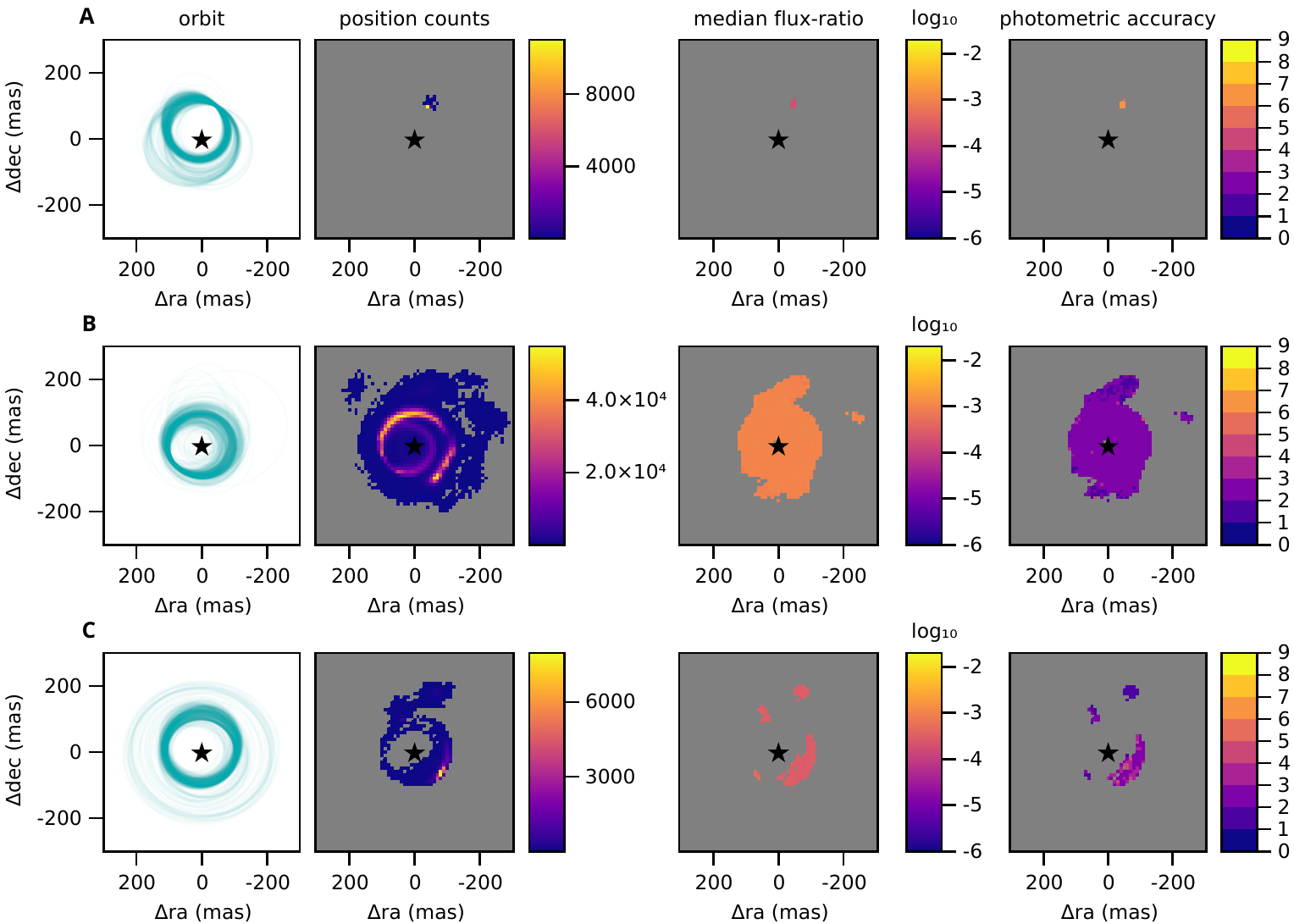}
    \caption{
    Comparison of the inner planet models against three different datasets.
    \textbf{A.} An inner planet model applied to the full dataset.
    \textbf{B.} Same as A, but ignoring the best epoch (2021).
    \textbf{C.} Same as A, but after subtracting the best fitting model found by A from the 2021 epoch.
    The four panels show orbits drawn from the posterior, posterior density of planet position calculated at the 2021 epoch, median photometry if the planet were at that position in 2021, and photometric accuracy at those positions i.e. the SNR of a planet if it were on an orbit that would pass through that pixel in 2021.
    Pixels are left blank where there are only a negligible number of samples.
    }\label{fig:inn-post-viz}
\end{figure*}

In Figure \ref{fig:inn-post-viz}, we draw orbits from the posterior and calculate their positions in 2021. We see that the model prefers a single location for the planet in 2021 North-North-West of the star. When we look at the median photometry of samples from the posterior that fall in this pixel, we find they are all roughly $2\times 10^{-4}$ in units relative contrast. Finally, when we look at the spread of the sample photometry, the ``photometric accuracy``, we find again a cluster of SNR 5-9.
These maps support the posterior being consistent with a single object rather than two or more.

Returning to the literature, 
{various candidate point sources have previously been reported. The candidate}
reported by \citet{maireLEECHExoplanetImaging2015} $3-4\sigma$ 0.2'' due South of the star in 2013, is not consistent with with our results. Likewise,
no compatible point sources are visible in the shorter wavelength (YJH) IFS data of \citet{wahhajz.SearchFifthPlanet2021}. Their IRDIS data in K band (closer to \Lprime) does show a low SNR point source North-West of the star in 2019; however, again, it's location may not be consistent with our orbital posterior to $>1\sigma$.
On the other hand,
the point source reported by \citet{currieDEEPTHERMALINFRARED2014}, $4\sigma$ North-North-West of the star in 2012, may have a roughly correct position angle if the candidate's semi-major axis is $\approx 4.7$, though with a slightly greater separation.
We did not include these sequences in our initial data selection (Section \ref{sec:obs-selection}), so to add them afterwards knowing it may or may not have a compatible point source could introduce confirmation bias.

This is an intriguing result,
but given the novelty of this analysis technique,
lack of confirmation at other wavelengths and instruments, 
and points discussed later in this analysis, we do not yet consider this a robust detection.

\begin{figure*}
    \centering
    \includegraphics[width=\textwidth]{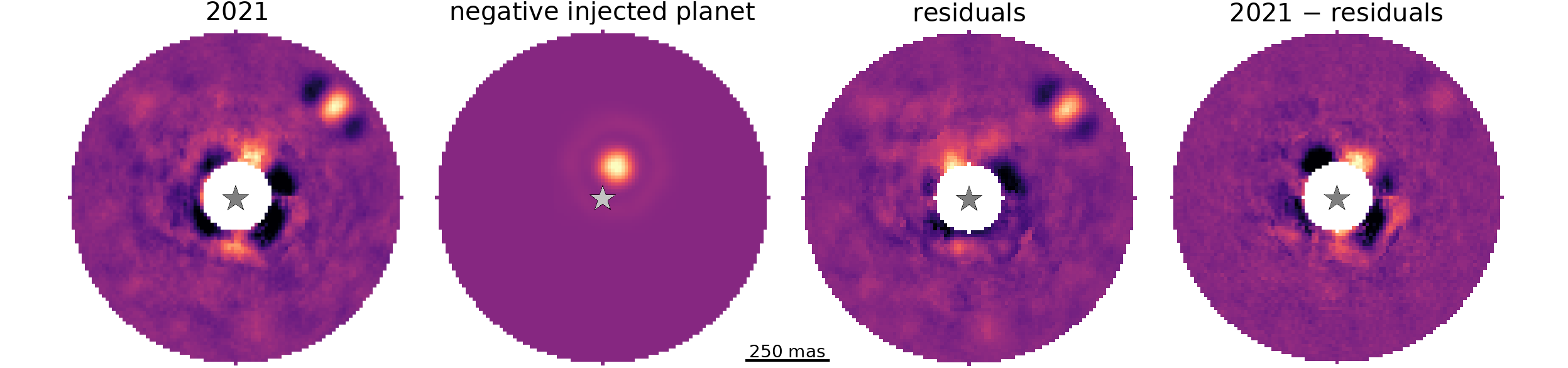}
    \caption{
        The 2021 epoch before and after subtracting a model planet injected at the expected location and photometry of the single inner planet model fit to all 5 epochs. The area interior to 100 mas is excluded from the processing due to detector saturation. {The bottom right image shows the flux removed by the negative planet injection. The structure is more complex than the typical dark wings from ADI processing due to the high FoV rotation and tight separation.}
    }\label{fig:planet-injection.png}
\end{figure*}

\subsection{Contribution of the 2021 Epoch}

Figure \ref{fig:inn-post-combined} shows that the preferred location of model is the most localized in 2021. This was our best epoch and consists of observations taken over four nights. We now examine the impact of this epoch on the model in greater detail.

To begin, we injected a negative model planet into the 2021 sequences raw data prior to SNAP reduction.
We placed the planet at the posterior expected position and photometry calculated from the full model of all five epochs.
We then re-reduced the data with SNAP to produce Figure \ref{fig:planet-injection.png}.

{The panels in that figure show the image before and after injecting the negative planet model}. The cente panel giving the difference between these reductions shows the flux removed by the planet model.
The model reproduces much of the irregular structure directly around the star including a bright lobe opposite to the expected position.
If the candidate is real, those effects can be understood as artifacts of the SNAP algorithm, tight separation, and rapid, near-$180\degree$ field of view rotation as the system transits the meridian from Maunakea.
In fact, the angle between the two bright spots is just under the average field of view rotation in the 2021 sequences.
Encouragingly, the opposite bright lobe is not picked up by the inner planet model meaning that a planet at that location and brightness in 2021 is not consistent with the other epochs.
That said, the injected planet model is not a perfect match for the candidate. The negative side-lobes (artifacts of any ADI reduction) are somewhat mismatched with the model having a darker sidelobe to the East than and lighter to the West than the candidate. As a consequence, some flux remains to the North-East of the star after subtraction.

Next, we also run our model on the original unmodified data but fully exclude the 2021 epoch. In this case, we again find no detection though interestingly the earlier four epochs still predict a spot of high posterior density within $\sim 1 \lambda/D$ of the location found in the full model.
These results do not mean that the earlier four epochs do not contribute to the SNR of the planet candidate.  They still contribute by ruling out large swaths of the orbital parameter space and the bright area South of the star in 2021. 

\begin{figure}[b]
    \centering
    \includegraphics{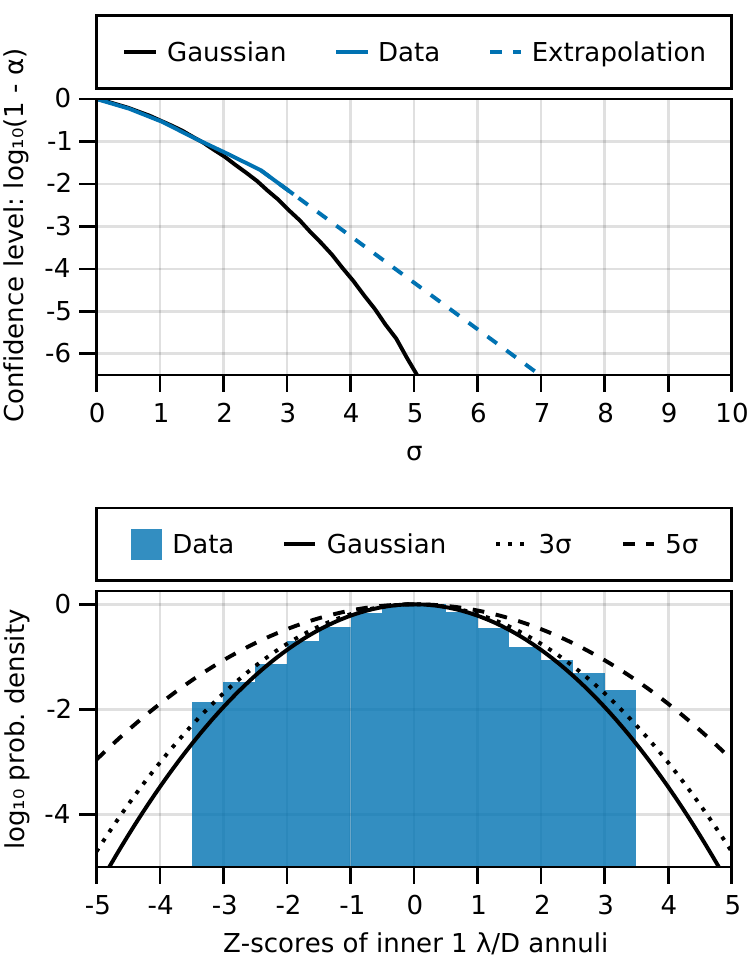}
    \caption{\textbf{Top:} Confidence levels of a Gaussian distribution and empirical confidence level  from the inner $1 \lambda/D$ annulus of our images extrapolated to $1-10^{-6}$ using an exponential function. We increase our SNR thresholds to account for this slight deviation from Gaussian noise. \textbf{Bottom:} Empirical log PDF of standardized pixel intensities compared to a Gaussian model, and Gaussians scaled to cover the tails of the distribution.}
    \label{fig:epdf}
\end{figure}

\newpage
\subsection{Noise Distribution and Sample Size}\label{sec:noise}

\begin{figure*}
    \centering
    \includegraphics[width=\textwidth]{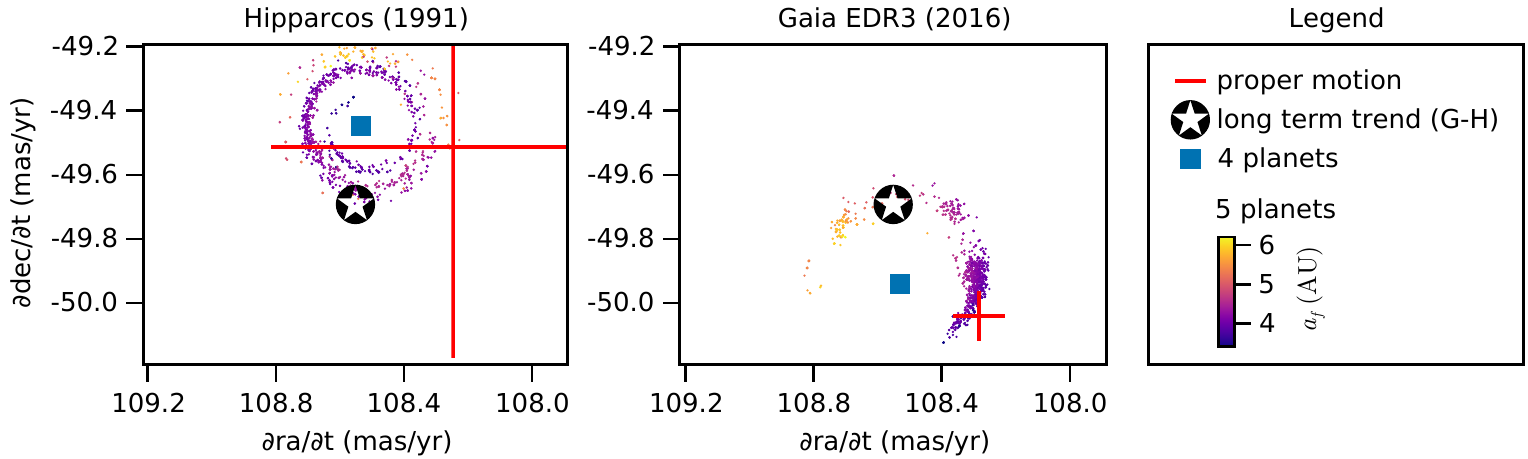}
    \caption{{Stellar astrometric motion predicted at the Hipparcos and GAIA epochs for the stable 4 planet $\mathrm{V_d}$ solution of \citet{gozdziewskiMultipleMeanMotion2014} (blue square) and with the addition of the fifth inner candidate (colored by the semi-major axis of the candidate). 
    The red error bars show the stellar astrometric motion at the Hipparcos and GAIA epochs as calculated by the HGCA \citep{brandtHipparcosGaiaCatalogAccelerations2021}.
    The black marker shows the long term proper motion from the HGCA calculated from the difference in position between both epochs.
    The planets c, d, e, and the candidate f are assumed to have masses of 7 \Mjup, while b is assumed to have a mass of 5 \Mjup. }
    }\label{fig:pma-comparison}
\end{figure*}

When evaluating a candidate in direct imaging, we should consider non-Gaussian noise and small sample statistics. Both act to increase the false positive fraction (FPF) and reduce our confidence in a detection. We calculate penalty factors for both of these effects on $3\sigma$ and $5\sigma$ Gaussian equivalent FPFs.

First, we consider non-Gaussian noise.
{Following \citet{maroisConfidenceLevelSensitivity2008}},
we compare the distribution of our data at the separation of the inner planet candidate to a Gaussian and estimate its effect on both detection thresholds.
In the top panel of Figure \ref{fig:epdf}, we plot the confidence levels of a Gaussian distribution and of standardized pixel data from the inner $1 \lambda/D$ annulus of all five backwards-rotated photometry maps. We extrapolate our data by fitting an exponential, and find that to reach a confidence level equivalent to a Gaussian at $3\sigma$ and $5\sigma$, we should penalize our SNR by factors of {1.07 and 1.35} respectively. 
The bottom panel of Figure \ref{fig:epdf} shows the log probability density function (PDF) of a Gaussian distribution and an empirical PDF (EPDF) of standardized pixel data from the inner $1 \lambda/D$ annulus of the backwards-rotated photometry maps. 
By expanding the Gaussian by these factors, it fully encompasses our data at a z-score of 3. The backwards rotated noise maps do not contain any pixels at z-scores beyond {3.5} (the forwards reduction does of course, since it contains the signal of the candidate) but the extrapolation appears valid and conservative.
These factors are relatively small indicating that the residual noise in each epoch is close to Gaussian distributed. This is not surprising due to the central limit theorem since, besides 2010, each epoch is a stack of 3-5 sequences with uncorrelated noise.

Next, we consider the effect of small sample statistics near the star. The contrast curves underlying our model at each epoch are calculated from as few as five independent samples at a separation of roughly $2 \lambda/D$.
If each epoch contributed to the SNR of our model in equal measure,
then we might consider the noise sample to be five times larger, reducing the effects of small sample statistics. However, the 2021 epoch contributes significantly to the overall figure. 
It's not {yet} clear how to correct this model for small sample statistics, but we can take a conservative approach by considering the final SNR to come only from {a single image (instead of 18 sequences)} and applying the correction factors of \citet{mawetFUNDAMENTALLIMITATIONSHIGH2014}, Table 1, at $2 \lambda/D$. This gives penalty factors of 1.35 and 2.2 for $3\sigma$ and $5\sigma$ FPFs respectively.

Combining these two sets of penalty factors, we should in fact apply $4.3\sigma$ and {$14.9\sigma$ thresholds} to reach large sample size, Gaussian equivalent FPFs for $3\sigma$ and $5\sigma$ respectively.
The candidate does easily meet the $3\sigma$ threshold but clearly falls well below a $5\sigma$ threshold for detection. This correction is perhaps overly pessimistic but reflects our goal to communicate this signal is a candidate worthy of additional study rather than an unambiguous detection.

\subsection{Mass and Proper Motion Anomaly}\label{sec:inn-mass}

{
The photometry of the inner planet candidate is consistent with planets c, d, and e, and brighter than b. Without photometry at other wavelengths, we therefore assume that it would have a similar mass to the inner planets c, d, and e.}

{The masses of the HR8799 planets have been estimated using several approaches.
From the beginning, luminosity modeling of the planets has suggested masses of approximately $7 \; \mathrm{M_{jup}}$ for the inner planets \citep{maroisDirectImagingMultiple2008a}.
Orbital stability analyses by \citet{wangDynamicalConstraintsHR2018} support the inner planets  having masses up to approximately $7 \; \mathrm{M_{jup}}$, though small islands of stability may exist for high masses.
\citet{sepulvedaDynamicalMassExoplanet2022} on the other hand, model the orbits of the planets in order to constrain the dynamical mass of the star. By combining this stellar mass with stellar and planet evolution models, they find that the inner planets c, d, and e likely have masses in the range of $4.1 - 7.0 \; \mathrm{M_{jup}}$ where age is the dominant contributor to the uncertainty.
In contrast to these estimates, \citet{brandtFirstDynamicalMass2021} find somewhat higher masses of $9.6^{+1.9}_{-1.8} \; \rm{M_{jup}}$ by combining the stable orbits found by \citet{wangDynamicalConstraintsHR2018} with the Hipparcos-Gaia Catalog of Accelerations \citep[HGCA,][]{brandtHipparcosGaiaCatalogAccelerations2021}
which calibrates the Hipparcos catalog against Gaia EDR3 \citep{gaia-collaborationGaiaEarlyData2021}.
The addition of a fifth massive planet would alter the solution space for all three methods incorporating orbital dynamics; however, the method based on proper motion anomaly would be the most impacted.
We now examine if this potential addition could account for the slight tension between these mass estimates.}

{Taking a more basic approach than that of \citet{brandtFirstDynamicalMass2021}, we model the proper motion anomaly of the star by assuming that the GAIA and Hipparcos missions each independently measure the position and instantaneous proper motion 25 times spaced equally throughout their respective missions.
We then compare these quantities against observations for a four planet and five planet model. 
For the four planet model, we use the stable orbital parameters of \citet{gozdziewskiMultipleMeanMotion2014} and fixed masses of  5 \Mjup\ and 7 \Mjup\ for b, and c, d, e respectively. For the five planet model, we use the same parameters but add a fifth inner planet from our orbital posterior with a mass of 7 \Mjup.}

{Figure \ref{fig:pma-comparison} shows the result of this comparison. Proper motion anomaly at the Hipparcos epoch is consistent with either the four or five planet model, but agreement at the GAIA epoch is significantly improved by the addition of the candidate at the most likely semi-major axis.}

{Though far from conclusive, this analysis demonstrates that the inclusion of the inner candidate could fully account for the observed proper motion anomaly if the candidate and inner planets c, d, and e have masses close to $7 \; \mathrm{M_{jup}}$.
This is slightly at odds with the \citet{brandtFirstDynamicalMass2021} result since they use the HGCA data to conclude that any additional inner planet more massive
than $\approx 6 \; \Mjup$ between 3 and 8 AU are unlikely.
The discrepancy could be due to their more sophisticated modeling or the lower masses used in our experiment for the known planets.}

\begin{figure*}
    \centering
    \includegraphics[width=0.49\textwidth]{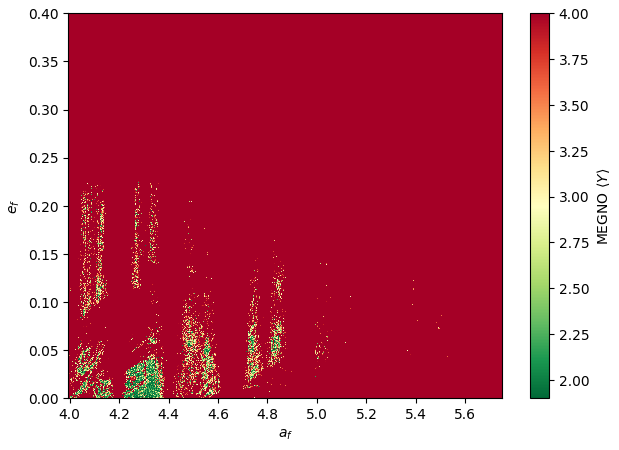}
    \includegraphics[width=0.49\textwidth]{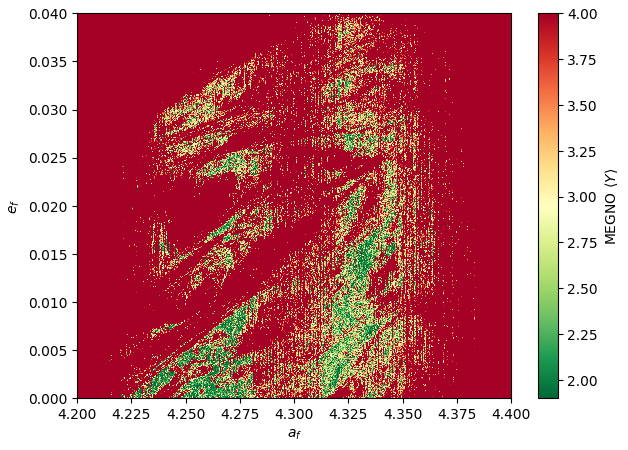}
    \caption{The MEGNO stability map for the coplanar `f' candidate using the parameters from Table \ref{tab:coplanarsol}, for an integration time of {0.5 Myr (left) and 1.0 Myr (right)}. Most of the configurations become unstable (MEGNO $>$ 2) within this integration period.}
    \label{fig:megno_05myr_coplanar}
\end{figure*}

\begin{deluxetable*}{c | r r r r r r r r}
\tablecaption{Stable orbits\label{tab:coplanarsol}}
\tablenum{4}
\tablehead{\colhead{Planet} & \colhead{Mass} & \colhead{a} & \colhead{e} & \colhead{i}& \colhead{$\Omega$}& \colhead{$\omega$}& \colhead{$\tau$} \\ 
\colhead{} & \colhead{($M_J$)} & \colhead{(AU)} & \colhead{} & \colhead{(\degree)} & \colhead{(\degree)} & \colhead{(\degree)} &\colhead{(\degree)} & \colhead{}} 
\startdata
b & 6.527020 & 68.597249 & 0.017425 & 27.502 & 63.953507 & 37.631852 &    0.872678 \\
c & 11.868008 &  39.486207 & 0.054102 & 27.502 & 63.953507 & 89.946387 &  0.403048\\
d & 7.178005 & 25.705213 & 0.137796 & 27.502 & 63.953507 & 33.186128 &  0.139618 \\
e &  6.298260 & 15.660910 & 0.168239 & 27.502 & 63.953507 & 110.074917 &  0.902362 \\
coplanar `f'  &  3.75000 & 4.325000 & 0.068600 & 27.502 & 63.953507 & 145.767982 &  0.833691 \\
non-coplanar `f'  &  3.75000 &  4.510300 & 0.036997 & 14.986 & 82.304917 & 119.041162 & 0.837036 \\
\enddata
\tablecomments{Stable five planet solutions. The non-coplanar solution is drawn from our orbital posterior and is stable for 0.75Myr. The coplanar solution was found using using a search grid near the orbital parameters of the posterior of the inner planet model. The stellar mass is 1.716162 $M_\odot$. The parameters of planets b, c, d, and e are from the \cite{gozdziewskiMultipleMeanMotion2014} $\mathrm{V_d}$ model.}
\end{deluxetable*}

\newpage
\subsection{Stability}

{We now consider how the addition of an inner planet would impact the stability of the system. For these tests we adopt a smaller than realistic mass for the candidate inner planet of $3.75\Mjup$ to ease the search for stable orbits.}
For our 5-planet models obtained entirely from the posteriors presented on Figure \ref{fig:model-comparison-all-up}, we found that no configuration is stable for the age of the system of 10-23 Myr \citep{sepulvedaDynamicalMassExoplanet2022}. When using the $\mathrm{V_d}$ Model parameters presented on \citet{gozdziewskiMultipleMeanMotion2014} for planets b, c, d and e we find {one unconstrained orbit for `f' that remains stable for 0.4 Myr.
A grid search around this sample's semi-major axis and eccentricity revealed similar orbits that are stable for up to 0.75Myr.
A grid search over co-planar orbits found regions that remain stable for up to 1.5 Myr. Most of the other configurations found from the posteriors become unstable within the first 0.5 Myr in our N-body simulations. These configurations of stable orbits are presented in Table \ref{tab:coplanarsol} and MEGNO stability maps of the coplanar orbits are presented in Figure \ref{fig:megno_05myr_coplanar}}.
{
These simulations do yet not explore changing the masses or orbital parameters of planets b, c, d, and e. It is likely that regions of greater stability could be found if these parameters were also explored in a future analysis.}

\subsection{Sensitivity to planets besides the 4-5 AU candidate}

\begin{figure}
    \centering
    \includegraphics[width=\colwidth]{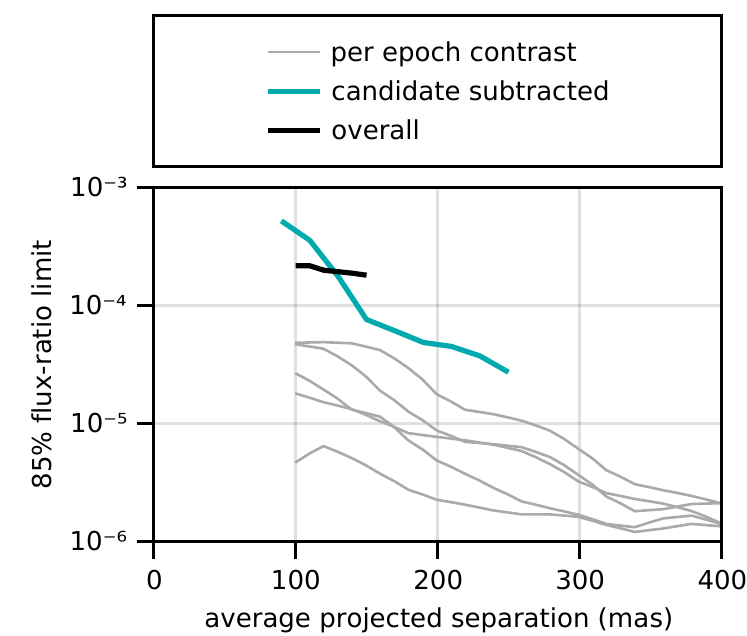}
    \caption{85th percentile upper limit of the marginal L band flux-ratio posterior for an additional fifth inner planet (black). The turquoise lines show the same, but after subtracting a planet model from the raw data of 2021 epoch (\textbf{C} in Figure \ref{fig:inn-post-viz}).
    {The gray lines show the traditional $1\sigma$ projected contrast of individual epochs from backwards rotated reductions}.
    }
    \label{fig:contrast-inn}
\end{figure}

Figure \ref{fig:contrast-inn} presents our sensitivity to additional inner planets besides the candidate near 4-5 AU. We show the 85th percentile of the marginal L flux-ratio posterior, conditioned on different ranges of semi-major axis. We show the sensitivity for both the regular dataset, and the dataset in which we subtracted the candidate from the 2021 epoch.
Ignoring the candidate presented above, assuming the system ages of \citet{sepulvedaDynamicalMassExoplanet2022} (10-23 Myr) and extrapolating the COND models of \citet{baraffeEvolutionaryModelsCool2003}, this translates to {$1\sigma$ upper mass limits of 4.3 \Mjup~and 3.0 \Mjup~for planets with orbits that have time-averaged projected separations of 150 mas and 250 mas respectively.
Note that this sensitivity versus average projected separation is not directly equivalent to contrast versus  separation at a given epoch as is usually quoted in the literature.}
In general, the sensitivity is a function of all orbital parameters. For instance, it improves with higher eccentricity at small semi-major axes since such a planet would spend more time away from the star.

\section{Conclusion}\label{sec:discussion}

In this paper we presented a deep targeted search in the HR 8799 system for additional planets using data at 3.8 microns.
\begin{itemize}
    \item We observed HR 8799 for 14 quarter nights with NIRC2 between 2017 and 2021
    \item We further gathered a further 4 quarter nights from the Keck archive for a 12 year baseline.
    \item We processed the data using direct S/N optimization \citep{thompsonImprovedContrastImages2021a} to improve contrasts at very small separations
    \item We used the Hamiltonian Monte Carlo method to explore both the space of possible orbits and flux from planets, allowing us to search for planets in the presence of orbital motion.
    This approach differs from the K-Stacker \citep{nowakKStackerKeplerianImage2018} technique since it requires the flux to be consistent across epochs, gives the covariance between the planet's flux and orbital parameters, and ultimately gives a detection that can be evaluated against uncertainties in both flux and orbital parameters.

\end{itemize}

Using these methods, we presented the results of our search:
\begin{itemize}
    \item We demonstrated this technique by re-detecting the four known planets b, c, d, and e at very high SNR 
    \item We presented limits on the flux of any additional outer planets between b and the start of the debris disk and did not find any additional outer planets above a $5 \sigma$ significance threshold, or a mass limit of roughly 0.9 \Mjup.
    \item We found a modest SNR candidate interior to the orbit of planet e. This planet would have a semi-major axis of approximately 4-5 AU, and L-band photometry similar to the inner planets c, d, and e. {According to the analysis \citet{sepulvedaDynamicalMassExoplanet2022} this would imply a mass of 4.1 to 7.0 \Mjup.}
\end{itemize}

For this candidate planet, we considered several factors.
\begin{itemize}
    \item We verified that subtracting a planet model from the raw data at the appropriate position and intensity in the best epoch and re-running the model removes the detection.
    \item We adjusted our detection thresholds to account for slightly non-Gaussian noise and a conservative treatment of the impacts of small sample statistics.
    \item {We showed that the addition of a fifth inner planet could improve the agreement between the dynamical mass measurements of the planets and mass estimates/limits from atmosphere models and orbital stability modeling.}
    \item {We performed rejection sampling with REBOUND using the inner candidate orbital posterior combined with a four planet solution known to be stable. We were successful in finding five planet orbital solutions that were stable for 0.75Myr using a lowered planet mass. A grid search over nearby co-planar orbital parameters found small families of orbits that are stable for up to $1.5\mathrm{Myr}$}.
\end{itemize}

Overall, we found that the inner candidate at SNR 6.9 easily met a $3\sigma$ equivalent FPF threshold ($4.3\sigma$) but does not meet a $5\sigma$ equivalent FPF threshold ($14.9\sigma$). This is primarily because of uncertainty in the contrast at each epoch due to the limited sample size at small separations from the star.
We consider this evidence intriguing, but caution that these results fall short of a conclusive detection.

{
\citet{wahhajImprovingSignaltonoiseDirect2015} presented the most sensitive limits on the K-band flux of an additional inner planet.
They found a $5\sigma$ upper limit of $3\times10^{-5}$ relative to the star at 100 mas separation. Adjusting this figure for small sample statistics in the same was as our data gives an upper limit of $7\times10^{-5}$.
As such, their non-detection of this candidate at K-band is a point against the candidate.
If on the other hand the candidate is confirmed,
this would give the candidate a very red color compared to c, d, and e 
($\mathrm{K-L} >2$).}
If real, such a red color might be caused by the different environment much closer to the star{.
One possible explanation might lie in photochemical hazes as proposed for 51 Eri b \citep{zahnlePHOTOLYTICHAZESATMOSPHERE2016,macintoshDiscoverySpectroscopyYoung2015} due to such a planet receiving more than an order of magnitude more light from the star than planet e.}

Additional follow up observations would be necessary to confirm this candidate; 
however, assuming an ideal $\sqrt{N}$ growth in SNR and a decreasing penalty for small sample statistics, this would require a further 12 quarter nights with NIRC2 of {similar quality to 2021}.
Instead, the best chance at confirming or rejecting this candidate may come from upcoming instruments with improved contrast at 100-250mas.
Followup observations from GRAVITY \citep{collaborationFirstLightGRAVITY2017} might be possible, but would be challenging given the remaining uncertainty in the candidate's orbit.
Another avenue that may be worth exploring is searching for the candidate with a fiber-fed spectrograph like {the Keck Planet Imager and Characterizer} \citep{delormeKeckPlanetImager2021}, though this again requires a well-determined orbit.

Regardless of if this candidate is confirmed, we demonstrated the utility of searching for planets in direct images by combining orbit modeling and planet detection.
This approach could considerably loosen scheduling requirements when searching for rapidly moving targets like planets around Alpha Cent, in addition to making the best use of direct imaging archives.

\begin{acknowledgments}
\section{Acknowledgements}

{The authors thank Dori Blakely, Thayne Curry, and the anonymous referee whose feedback have greatly improved this work.
WT acknowledges the support of the Natural Sciences and Engineering Research Council of Canada (NSERC), 466479467.}
This research used the facilities of the Canadian Astronomy Data Centre operated by the National Research Council of Canada with the support of the Canadian Space Agency. 
The data presented herein were obtained at the W. M. Keck Observatory, which is operated as a scientific partnership among the California Institute of Technology, the University of California and the National Aeronautics and Space Administration. The Observatory was made possible by the generous financial support of the W. M. Keck Foundation.
This research has made use of the Keck Observatory Archive (KOA), which is operated by the W. M. Keck Observatory and the NASA Exoplanet Science Institute (NExScI), under contract with the National Aeronautics and Space Administration.
This work has made use of data from the European Space
Agency (ESA) mission Gaia (https://www.cosmos.esa.int/
Gaia), processed by the Gaia Data Processing and Analysis
Consortium (DPAC, https://www.cosmos.esa.int/web/Gaia/
dpac/consortium). Funding for the DPAC has been provided
by national institutions, in particular the institutions participating in the Gaia Multilateral Agreement.

The authors wish to recognize and acknowledge the very significant cultural role and reverence that the summit of Maunakea has always had within the indigenous Hawaiian community.  We are most fortunate to have the opportunity to conduct observations from this mountain.

\end{acknowledgments}

\software{Julia \citep{bezansonJuliaFastDynamic2012},
AdvancedHMC.jl \citep{xuAdvancedHMCJlRobust2020},
REBOUND \citep{2015MNRAS.452..376R},
Makie.jl \citep{danischMakieJlFlexible2021},
Matplotlib \citep{hunterMatplotlib2DGraphics2007},
GR \citep{GRFramework}, PairPlots.jl.}

\medskip

\bibliography{manual.bib}

\appendix
\section{Additional Figures}\label{sec:adnl-images}

\begin{figure*}[!ht]
    \centering
        \includegraphics[width=\textwidth]{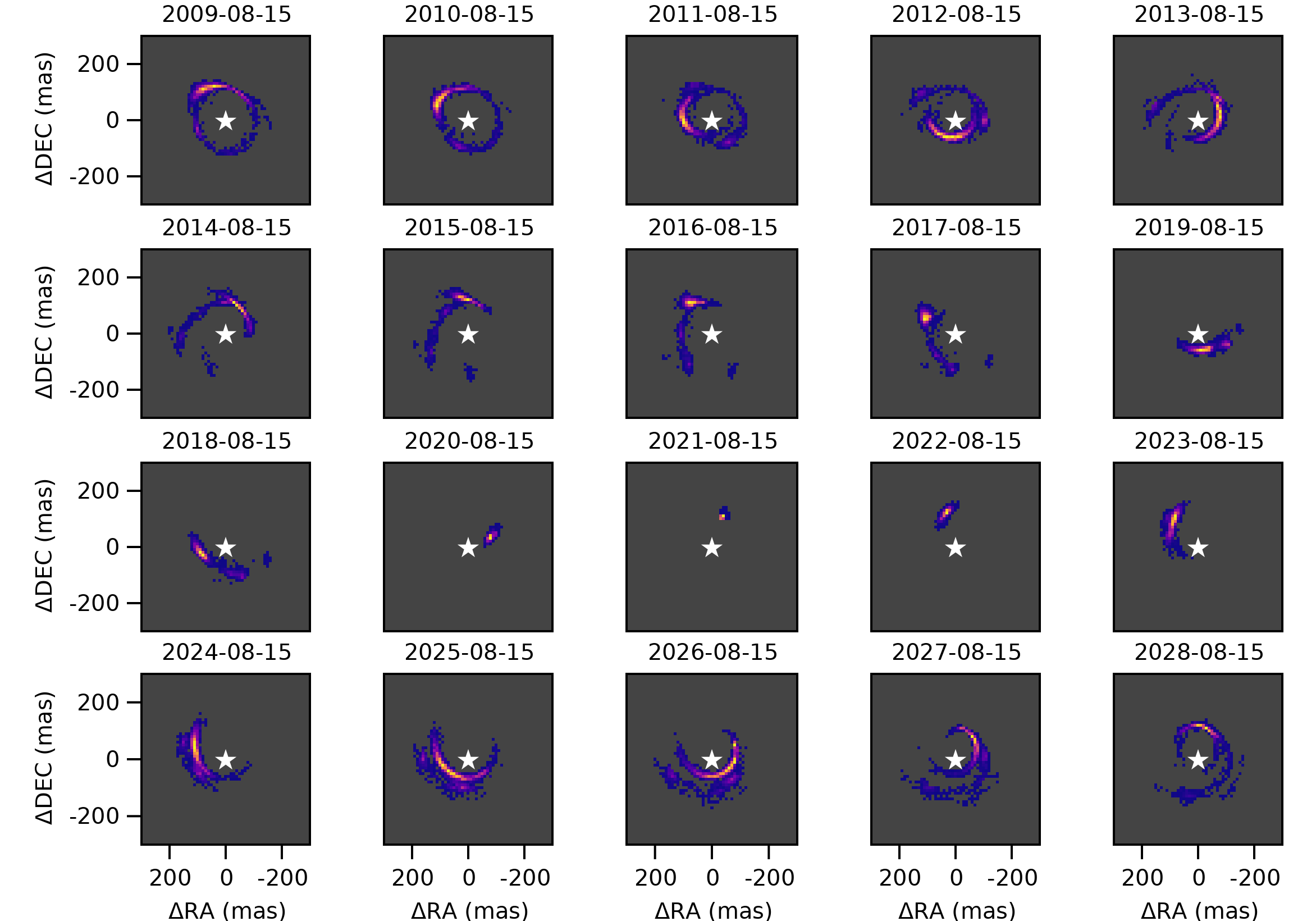}
    \caption{Position posterior density for the inner planet model at different epochs. The position is shown on August 15th of each year between 2009 and 2024.\label{fig:pos-posts}}
\end{figure*}

\begin{figure*}[!ht]
    \centering
        \includegraphics[width=\textwidth]{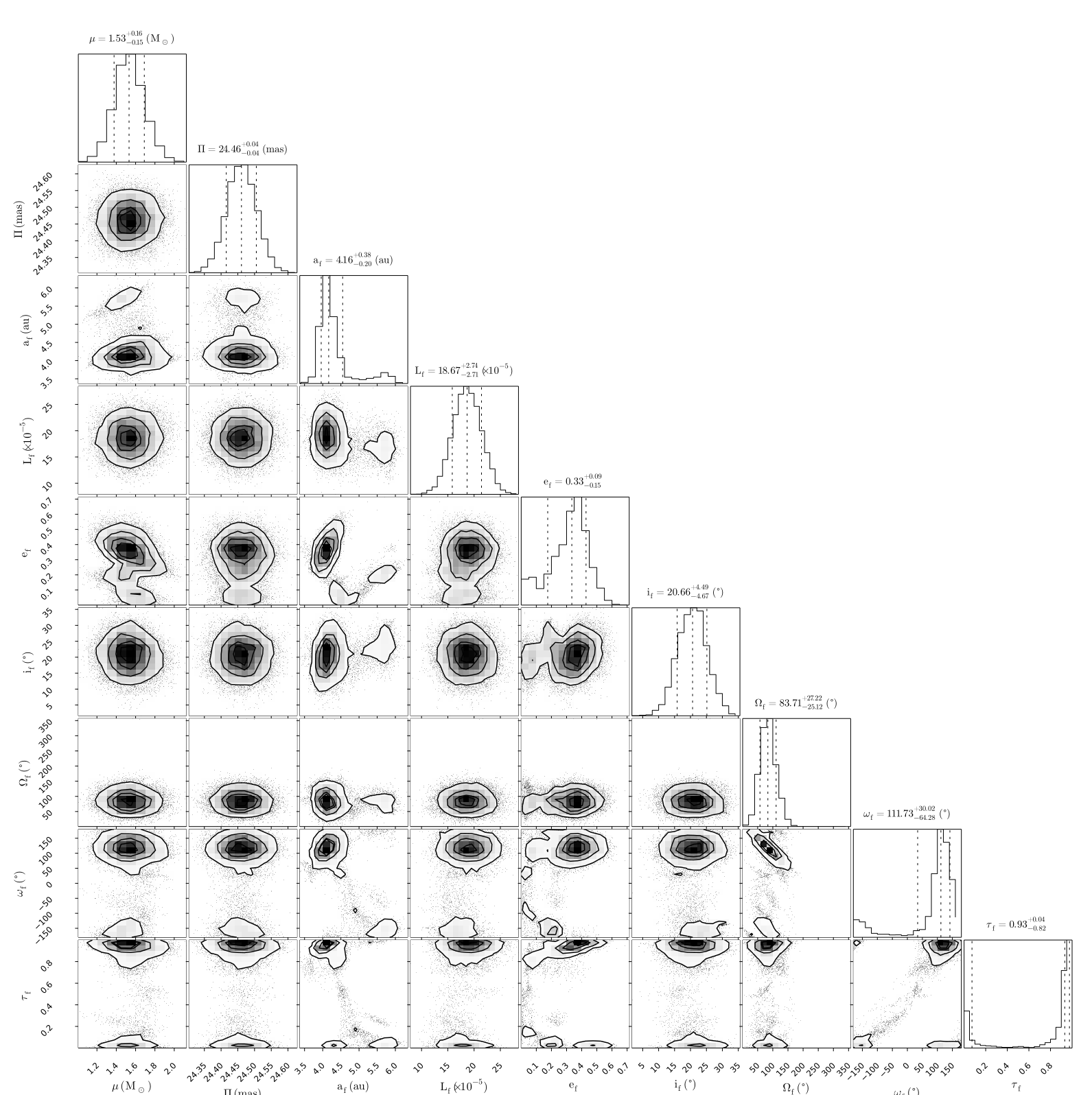}
    \caption{Corner plot showing the posterior of the inner planet model applied to all epochs. {Note that the angular parameters $i$, $\Omega$, $\omega$, and $\tau$ are periodic.} Chains are available at 10.5281/zenodo.6823071.}\label{fig:corner-f}
\end{figure*}

\begin{figure*}
    \centering
        \includegraphics[width=\textwidth]{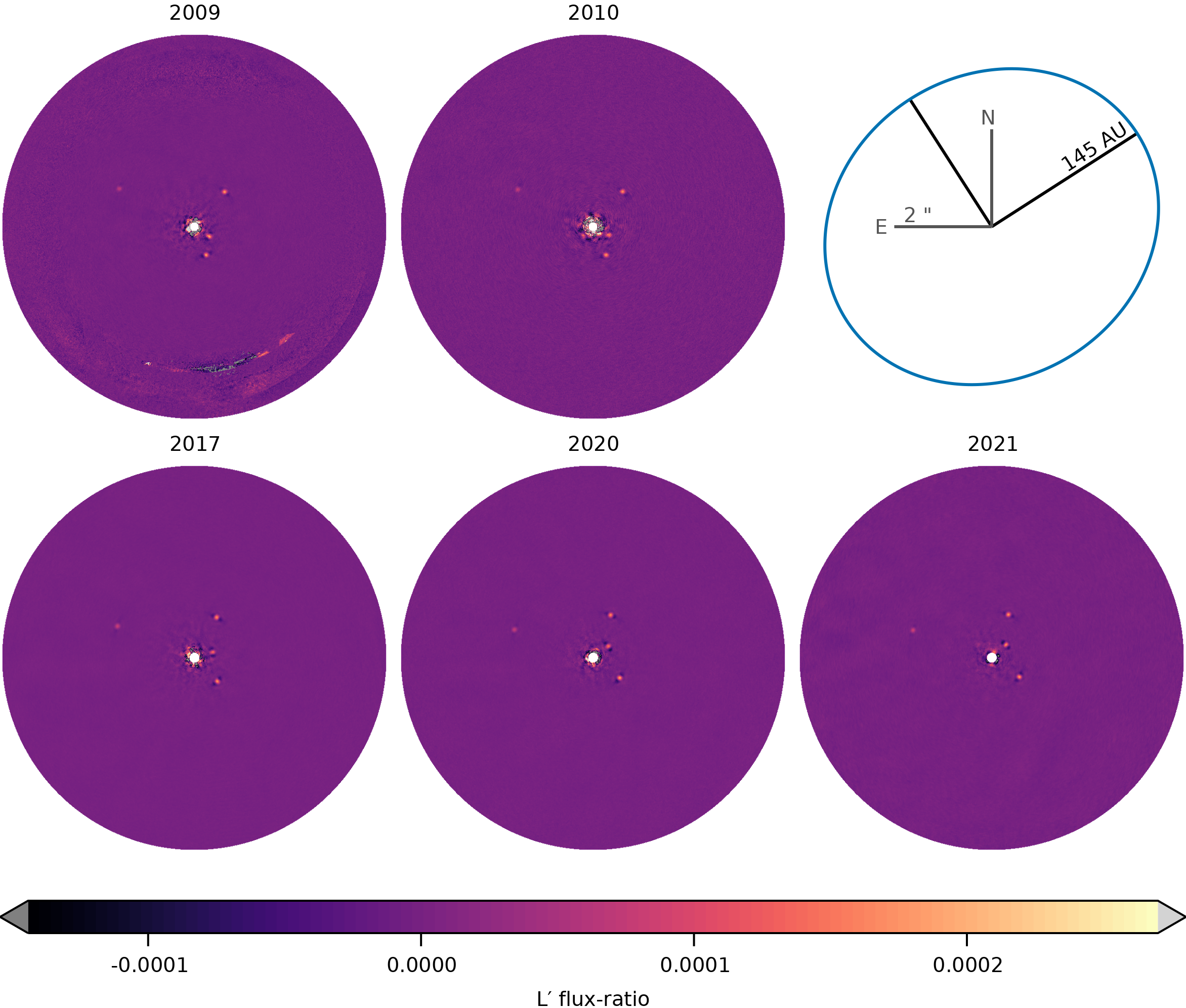}
    
    \caption{Full combined image from each epoch. The top two panels consist of archival data while the bottom three are from the new campaign. The artifacts south of the star in the 2009 image are from a chopping strategy used to subtract the thermal background. The approximate location of the inner edge of the outer debris disk is outlined in blue.
    These processed images are available at 10.5281/zenodo.6823071.
    }
\end{figure*}

\end{document}